\documentclass{IEEEtran}
\usepackage{amsmath,amssymb,amsthm,bm,cite,graphicx,multirow,tikz,verbatim}
\usepackage{microtype}
\usepackage[algo2e]{algorithm2e}

\fussy

\usetikzlibrary{arrows,chains,decorations,matrix,positioning,scopes,shadows,shapes,shapes.gates.logic.US,topaths}

\tikzset{
  nonterminal/.style={
    rectangle,
    minimum size=6mm,
    very thick,
    draw=red!50!black!50,         
    top color=white,              
    bottom color=red!50!black!20, 
    font=\itshape
  },
  terminal/.style={
    rounded rectangle,
    minimum size=6mm,
    very thick,draw=black!50,
    top color=white,bottom color=black!20,
    font=\ttfamily},
  skip loop/.style={to path={-- ++(0,#1) -| (\tikztotarget)}},
  hvpath/.style={to path={-| (\tikztotarget)}},
  vhpath/.style={to path={|- (\tikztotarget)}},
  tip/.style={->, shorten >=0.007pt},
  every join/.style={rounded corners}
}

{
  \tikzset{terminal/.append style={text height=1.5ex,text depth=.25ex}}
  \tikzset{nonterminal/.append style={text height=1.5ex,text depth=.25ex}}
}

\tikzstyle{point}=[coordinate]
\tikzstyle{conn}=[circle, fill, minimum size=3pt, inner sep=0pt, outer sep=0pt]
\tikzstyle{reg}=[rectangle, draw, solid, minimum size=12pt, inner sep=0pt, outer sep=0pt]
\tikzstyle{op}=[circle, draw, solid, minimum size=12pt, inner sep=0pt, outer sep=0pt]
\tikzstyle{mux}=[trapezium, draw, solid, shape border rotate=180, minimum size=12pt]
\tikzstyle{smux}=[trapezium, draw, solid, shape border rotate=0, minimum size=12pt]
\tikzstyle{emux}=[trapezium, draw, solid, shape border rotate=90, minimum size=12pt]
\tikzstyle{wmux}=[trapezium, draw, solid, shape border rotate=270, minimum size=12pt]
\tikzstyle{wmuxb}=[trapezium, draw, solid, shape border rotate=270, minimum size=14pt]
\tikzstyle{pe}=[rectangle, draw, solid, minimum size=34pt, inner sep=0pt, outer sep=0pt]
\tikzstyle{peb}=[rectangle, draw, solid, minimum size=40pt, inner sep=0pt, outer sep=0pt]
\tikzstyle{peg}=[rectangle, draw, solid, minimum size=47pt, inner sep=0pt, outer sep=0pt]

\newcommand{\SI}{}
\newcommand{\si}{}

\newcommand{\mm}{$\mathrm{mm}$}
\newcommand{\nm}{$\mathrm{nm}$}
\newcommand{\ns}{$\mathrm{ns}$}
\newcommand{\mega}{$\mathrm{M}$}
\newcommand{\bit}{$\mathrm{bit}$}
\newcommand{\per}{$\mathrm{/}$}
\newcommand{\second}{$\mathrm{s}$}
\newcommand{\squared}{$^2$}
\newcommand{\milli}{$\mathrm{m}$}
\newcommand{\watt}{$\mathrm{W}$}

\theoremstyle{definition}
\newtheorem{algorithm}{Algorithm}
\newtheorem{lemma}{Lemma}

\DeclareMathOperator*{\argmin}{arg\,min}
\DeclareMathOperator*{\rank}{rank}
\DeclareMathOperator*{\RRE}{RRE}
\DeclareMathOperator*{\minpoly}{minpoly}
\DeclareMathOperator*{\shiftup}{shiftup}
\DeclareMathOperator*{\shiftleft}{shiftleft}
\DeclareMathOperator*{\eliminate}{eliminate}
\DeclareMathOperator*{\reduce}{reduce}
\DeclareMathOperator*{\AAE}{AE}
\DeclareMathOperator*{\ACtrl}{ACtrl}
\DeclareMathOperator*{\BCtrl}{BCtrl}
\DeclareMathOperator*{\BE}{BE}
\DeclareMathOperator*{\GE}{GE}
\DeclareMathOperator*{\ME}{ME}
\DeclareMathOperator*{\QCtrl}{QCtrl}
\DeclareMathOperator*{\QE}{QE}
\DeclareMathOperator*{\ct}{ct}
\DeclareMathOperator*{\ctau}{ctau}
\DeclareMathOperator*{\ctal}{ctal}
\DeclareMathOperator*{\ctq}{ctq}
\DeclareMathOperator*{\crr}{cr}
\providecommand{\abs}[1]{\lvert#1\rvert}

\begin{document}
\title{Rank Metric Decoder Architectures for Random Linear Network Coding with Error Control}
\author{
Ning~Chen~\IEEEmembership{Member,~IEEE}, Zhiyuan~Yan~\IEEEmembership{Senior~Member,~IEEE}, Maximilien~Gadouleau~\IEEEmembership{Member,~IEEE}, Ying~Wang~\IEEEmembership{Member,~IEEE}, and Bruce~W.~Suter~\IEEEmembership{Senior~Member,~IEEE}
\thanks{This work was supported in part by Thales Communications, Inc.,  a summer extension grant from Air Force Research Lab, and NSF under grant ECCS-0925890.
The material in this paper was presented in part at the IEEE Workshop on Signal Processing Systems, Tampere, Finland, October 2009.}
\thanks{Ning Chen was with the Department of Electrical and Computer Engineering, Lehigh University, Bethlehem, PA 18015 USA.
Now he is with the Enterprise Storage Division, PMC-Sierra Inc., Allentown, PA 18104 USA (e-mail: ning\_chen@pmc-sierra.com).}
\thanks{Zhiyuan Yan is with the Department of Electrical and Computer Engineering, Lehigh University, Bethlehem, PA 18015 USA (e-mail: yan@lehigh.edu).}
\thanks{Maximilien Gadouleau was with the Department of Electrical and Computer Engineering, Lehigh University, Bethlehem, PA 18015 USA.
Now he is with the Department of Computer Science, Queen Mary, University of London, E1 4NS UK (e-mail: mgadouleau@eecs.qmul.ac.uk).}
\thanks{Ying Wang is with Qualcomm Flarion Technologies, Bridgewater, NJ 08807 USA (e-mail: aywang11@gmail.com).}
\thanks{Bruce W. Suter is with Air Force Research Laboratory, Rome, New York 13441 USA (e-mail: bruce.suter@rl.af.mil).}
}
\maketitle
\begin{abstract}
While random linear network coding is a powerful tool for disseminating information in communication networks, it is highly susceptible to errors caused by various sources. Due to error propagation, errors greatly deteriorate the throughput of network coding and seriously undermine both reliability and security of data. Hence error control for network coding is vital. Recently, constant-dimension codes (CDCs), especially K\"otter--Kschischang (KK) codes, have been proposed for error control in random linear network coding. KK codes can also be constructed from Gabidulin codes, an important class of rank metric codes.
Rank metric decoders have been recently proposed for both Gabidulin and KK codes, but they have high computational complexities. Furthermore, it is not clear whether such decoders are feasible and suitable for hardware implementations. In this paper, we reduce the complexities of rank metric decoders and propose novel decoder architectures for both codes. The synthesis results of our decoder architectures for Gabidulin and KK codes with limited error-correcting capabilities over small fields show that our architectures not only are affordable, but also achieve high throughput.
\end{abstract}

\section{Introduction}
Network coding \cite{Ahlswede00} is a promising candidate for a new unifying design paradigm for communication networks, due to its advantages in throughput and robustness to network failures. Hence, network coding is already used or considered in gossip-based data dissemination, 802.11 wireless ad hoc networking, peer-to-peer networks, and mobile ad hoc networks (MANETs).

Random linear network coding (RLNC) \cite{Ho06} is arguably the most important class of network coding.
RLNC treats all packets as vectors over some finite field and forms an outgoing packet by linearly combining incoming packets using random coefficients. Due to its random linear operations, RLNC not only achieves network capacity in a distributed manner, but also provides robustness to changing network conditions. Unfortunately, it is highly susceptible to errors caused by various reasons, such as noise, malicious or malfunctioning nodes, or insufficient min-cut~\cite{Kotter08}. Since linearly combining packets results in error propagation, errors greatly deteriorate the throughput of network coding and seriously undermine both reliability and security of data. Thus, error control for random linear network coding is critical.

Error control schemes proposed for RLNC assume two types of transmission models. The schemes of the first type (see, for example, \cite{Cai02}) depend on and take advantage of the underlying network topology or the particular linear network coding operations performed at various network nodes. The schemes of the second type \cite{Kotter08,Silva08}  assume that the transmitter and receiver have no knowledge of such channel transfer characteristics. The two transmission models are referred to as coherent and noncoherent network coding, respectively.

It has been recently shown \cite{Kotter08} that an error control code for noncoherent network coding, called a subspace code, is a set of subspaces (of a vector space), and information is encoded in the choice of a subspace as a codeword; a set of packets that generate the chosen subspace is then transmitted  \cite{Kotter08}.
A subspace code is called a \emph{constant-dimension code} (CDC) if its subspaces are of the same dimension.
CDCs are of particular interest since they lead to simplified network protocols due to the fixed dimension. A class of asymptotically optimal CDCs have been proposed in \cite{Kotter08}, and they are referred to as the KK codes. A decoding algorithm based on interpolation for bivariate linearized polynomials is also proposed in~\cite{Kotter08} for the KK codes. It was shown that KK codes correspond to \emph{lifting}~\cite{Silva08} of Gabidulin codes~\cite{Gabidulin85,Roth91}, a class of optimal rank metric codes.
Gabidulin codes are also called maximum rank distance (MRD) codes, since they achieve the Singleton bound in the rank metric~\cite{Gabidulin85}, as Reed--Solomon (RS) codes achieve the Singleton bound of Hamming distance.
Due to the connection between Gabidulin and KK codes, the decoding of KK codes can be viewed as generalized decoding of Gabidulin codes, which involves \emph{deviations} as well as errors and erasures~\cite{Silva08}.
Gabidulin codes are significant in themselves: For coherent network coding, the error correction capability of error control schemes is succinctly described by the \emph{rank metric} \cite{Silva09}; thus error control codes for coherent network coding are essentially rank metric codes.

The benefits of network coding above come at the price of additional operations needed at the source nodes for encoding, at the intermediate nodes for linear combining,
and at the destination node(s) for decoding. In practice, the  decoding complexities  at destination nodes are much greater than the encoding and combining complexities.  The decoding complexities of RLNC are particularly high when large underlying fields are assumed and when additional mechanisms such as error control are accounted for. Clearly, the decoding complexities of RLNC are critical to both software and hardware implementations. Furthermore, area/power overheads of their VLSI implementations are important factors in system design. Unfortunately, prior research efforts have mostly focused on theoretical aspects of network coding, and complexity reduction and efficient VLSI implementation of network coding decoders have not been sufficiently investigated so far. For example, although the decoding complexities of Gabidulin and KK codes were analyzed in~\cite{Gadouleau08a,Kschischang09} and \cite{Kotter08,Silva08}, respectively, they do not reflect the impact of the size of the underlying finite fields. To ensure high probability of success for RLNC, a field of size $2^8$ or $2^{16}$ is desired \cite{Chou03}. However, these large field sizes will increase decoding complexities and hence complicate hardware implementations.
Finally, to the best of our knowledge, hardware architectures for these decoders have not been investigated in the open literature.

In this paper, we fill this significant gap by investigating complexity reduction and efficient hardware implementation for decoders in RLNC with error control.
This effort is significant to the evaluation and design of network coding for several reasons. First, our results evaluate the complexities of decoders for RLNC as well as the area, power, and throughput of their hardware implementations, thereby helping to determine the feasibility and suitability of network coding for various applications. Second, our research results provide instrumental guidelines to the design of network coding from the perspective of complexity as well as hardware implementation. Third, our research results lead to efficient decoders and hence reduce the area and power overheads of network coding.

In this paper, we focus on the generalized Gabidulin decoding algorithm \cite{Silva08} for the KK codes and the modified Berlekamp--Massey decoding algorithm in \cite{Richter04} for Gabidulin codes for two reasons. First, compared with the decoding algorithm in~\cite{Kotter08}, the generalized Gabidulin decoding~\cite{Silva08} has a smaller complexity, especially for high-rate KK codes~\cite{Silva08}.
Second, components in the errors-only Gabidulin decoding algorithm in \cite{Richter04}
can be easily adapted in the generalized Gabidulin decoding of KK codes. Thus, among the decoding algorithms for Gabidulin codes, we focus on the decoding algorithm in \cite{Richter04}.

Although we focus on RLNC with error control in this paper, our results can be easily applied to RLNC without error control. For RLNC without error control, the decoding complexity is primarily due to inverting of the global coding matrix via Gauss-Jordan elimination, which is also considered in this paper.

Our main contributions include several algorithmic reformulations that reduce the computational complexities of decoders
for both Gabidulin and KK codes. Our complexity-saving algorithmic reformulations are:
\begin{itemize}
	\item We first adopt normal basis representations for all finite field elements, and then significantly reduce the complexity of bit-parallel normal basis multipliers by using our common subexpression elimination (CSE) algorithm \cite{Chen09};
	\item The decoding algorithms of both Gabidulin and KK codes involve solving key equations.
		We adapt the inversionless Berlekamp--Massey algorithm (BMA) in \cite{Burton71,Sarwate01} to solving key equations for rank metric codes.
		Our inversionless BMA leads to reduced complexities as well as efficient architectures;
	\item The decoding algorithm of KK codes requires that the input be arranged in a row reduced echelon (RRE) form \cite{Lancaster85}. We define a more generalized form called $n$-RRE form, and show that it is sufficient if the input is in the $n$-RRE form. This change not only reduces the complexity of reformulating the input, but also enables parallel processing of decoding KK codes based on Cartesian products.
\end{itemize}

Another main contribution of this paper is efficient decoder architectures
for both Gabidulin and KK codes. Aiming to reduce the area and to improve the regularity of our decoder architectures, we have also reformulated other steps in the decoding algorithm.
To evaluate the performance of our decoder architectures for Gabidulin and KK codes, we implement our decoder architecture for two rate-$1/2$ Gabidulin codes and their corresponding KK codes.
Our KK decoders can be used in network coding with various packet lengths by Cartesian product~\cite{Silva08}.
The synthesis results of our decoders show that our decoder architectures for Gabidulin and KK codes over small fields with limited error-correcting capabilities not only are affordable, but also achieve high throughput.
Our decoder architectures and implementation results are novel to the best of our knowledge.

The decoders considered in this work are bounded distance decoders, and their decoding capability is characterized in \cite[Theorem~11]{Silva08}.
The thrust of our work is to reduce complexities and to devise efficient architectures for such decoders, while maintaining their decoder capability.
To this end, our reformulations of the decoding algorithms do not affect the decoding capability of the bounded distance decoders of Gabidulin and KK codes.
The error performance of the bounded distance decoders has been investigated in our previous works \cite{Gadouleau08b,Gadouleau09a,Gadouleau09b}. Hence, despite its significance, a detailed error performance analysis is out of the scope of this paper, and we do not include it due to limited space.

The rest of the paper is organized as follows. After briefly reviewing the background in Section~\ref{sec:pre}, we present our complexity-saving algorithmic reformulations  and efficient decoder architectures in Sections~\ref{sec:com} and \ref{sec:arch}, respectively.
In Section~\ref{sec:impl}, the proposed architectures are implemented in Verilog and synthesized for area/performance evaluation. The conclusion is given in Section~\ref{sec:con}.

\section{Preliminaries}\label{sec:pre}

\subsection{Notation}
Let $q$ denote a power of prime and $\mathbb{F}_{q^m}$ denote a finite field of order $q^m$.
We use  $\bm{I}_n$, $\mathbb{F}_q^{n}$, and $\mathbb{F}_q^{n\times m}$ to denote an $n\times n$ identity matrix, an $n$-dimensional vector space over $\mathbb{F}_q$, and the set of all $n \times m$ matrices over $\mathbb{F}_q$, respectively.
For a set $\mathcal{U} \subseteq \{0,1,\dotsc,n-1\}$, $\mathcal{U}^c$ denotes the complement subset $\{0,1,\dotsc,n-1\} \setminus \mathcal{U}$ and $\bm{I}_\mathcal{U}$ denotes the columns of $\bm{I}_n$ in $\mathcal{U}$.
In this paper, all vectors and matrices are in bold face.

The rank weight of a vector over $\mathbb{F}_{q^m}$ is defined as the \emph{maximal} number of its coordinates that are linearly independent over the base field $\mathbb{F}_q$.
Rank metric between two vectors over $\mathbb{F}_{q^m}$ is the rank weight of their difference~\cite{Delsarte78}.
For a column vector $\bm{X} \in \mathbb{F}_{q^m}^n$, we can expand each of its component into a row vector over the base field $\mathbb{F}_q$.
Such a row expansion leads to an $n\times m$ matrix over $\mathbb{F}_q$.
In this paper, we slightly abuse the notation so that $\bm{X}$ can represent a vector in $\mathbb{F}_{q^m}^n$ or a matrix in $\mathbb{F}_q^{n\times m}$, although the meaning is usually clear given the context.

Given a matrix $\bm{X}$, its row space, rank, and reduced row echelon (RRE) form are denoted by $\langle\bm{X}\rangle$, $\rank\bm{X}$, and $\RRE(\bm{X})$, respectively.
For a subspace $\langle \bm{X}\rangle$, its dimension is denoted by $\dim\langle\bm{X}\rangle$ and $\rank\bm{X}=\dim\langle\bm{X}\rangle$.
The rank distance of two vectors $\bm{X}$ and $\bm{Y}$ in $\mathbb{F}_{q^m}^n$ is defined as $d_R(\bm{X}, \bm{Y}) \triangleq \rank(\bm{X}-\bm{Y})$.
The subspace distance~\cite{Kotter08} of their row spaces $\langle\bm{X}\rangle, \langle\bm{Y}\rangle$ is defined as $d_S(\langle\bm{X}\rangle, \langle\bm{Y}\rangle) \triangleq \dim\langle\bm{X}\rangle + \dim\langle\bm{Y}\rangle - 2\dim(\langle\bm{X}\rangle \cap \langle\bm{Y}\rangle)$.

A \emph{linearized polynomial}~\cite{Ore33,Ore34} (or $q$-polynomial) over $\mathbb{F}_{q^m}$ is a polynomial of the form $f(x) = \sum_{i=0}^p f_i x^{q^i}$, where $f_i \in \mathbb{F}_{q^m}$. For a linearized polynomial $f(x)$,  its $q$-degree is defined to be the greatest value of $i$ for which $f_i$ is non-zero.  For convenience, let $[i]$ denote $q^i$.
The symbolic product of two linearized polynomials $a(x)$ and $b(x)$, denoted by $\otimes$ (that is, $a(x)\otimes b(x)= a(b(x))$),
is also a linearized polynomial.
The $q$-reverse of a linearized polynomial $f(x) = \sum_{i=0}^p f_i x^{[i]}$ is given by the polynomial $\bar{f}(x) = \sum_{i=0}^p \bar{f}_i x^{[i]}$, where $\bar{f}_i = f_{p-i}^{[i-p]}$ for $i=0,1,\dotsc,p$ and $p$ is the $q$-degree of $f(x)$.
For a set $\bm{\alpha}$ of field elements, we use $\minpoly(\bm{\alpha})$ to denote its \emph{minimal linearized polynomial}, which is the monic linearized polynomial of least degree such that all the elements of $\bm{\alpha}$ are its roots.

\subsection{Gabidulin Codes and Their Decoding}
A Gabidulin code~\cite{Gabidulin85} is a linear $(n, k)$ code over $\mathbb{F}_{q^m}$, whose parity-check matrix has a form as
\begin{equation} \bm{H} = \begin{bmatrix} h_0^{[0]} & h_1^{[0]} & \dotsb & h_{n-1}^{[0]}\\
	h_0^{[1]} & h_1^{[1]} & \dotsb & h_{n-1}^{[1]}\\
	\vdots & \vdots & \ddots & \vdots\\
	h_0^{[n-k-1]} & h_1^{[n-k-1]} & \dotsb & h_{n-1}^{[n-k-1]}
\end{bmatrix}\label{eqn:check}
\end{equation}
where $h_0, h_1, \dotsc, h_{n-1} \in \mathbb{F}_{q^m}$ are linearly independent over $\mathbb{F}_q$.
Let $\bm{h}$ denote $(h_0, h_1, \dotsc, h_{n-1})^T$.
Since $\mathbb{F}_{q^m}$ is an $m$-dimensional vector space over $\mathbb{F}_q$, it is necessary that $n \le m$.
The minimum rank distance of a Gabidulin code is $d = n - k + 1$, and hence Gabidulin codes are MRD codes.

\begin{figure}[htbp]
	\centering
\begin{tikzpicture}
\ifCLASSOPTIONonecolumn
	\node{
\else
	\node[scale=0.65]{
\fi
	\begin{tikzpicture}[draw=black!50]
    \matrix[row sep=6mm, column sep=4mm] {
        \node (words) [nonterminal] {Received};&
        \node (p2) [conn] {}; & \node (synds) [terminal]    {Syndromes};&
        \node (p3) [conn] {}; & \node (bma) [terminal]    {BMA};&
        \node (p4) [point] {};\\
        \node (codes) [nonterminal] {Corrected};&
		\node [point] {}; &
        \node (es) [terminal] {Error}; &
		\node (p7) [point] {};&
		\node (gabidulin) [terminal] {Gabidulin's};&
		\node (p5) [conn] {};&
		 \node (rs)   [terminal]    {Roots};\\
    };

	\node[above of=p2, yshift=-20pt] {$\bm{r}$};
	\node[above of=p3, yshift=-20pt] {$\bm{S}$};
	\node[above of=p4, yshift=-20pt] {$\sigma(x)$};
	\node[above of=p5, yshift=-20pt] {$\bm{E}$};
	\node[above of=p7, yshift=-20pt] {$\bm{X}$};
    { [start chain]
        \chainin (words);
        \chainin (synds)    [join=by tip];
        \chainin (bma)   [join=by tip];
		\chainin (rs)    [join=by {hvpath, tip}];
		\chainin (gabidulin) [join=by tip,join=with p3 by {skip loop=-6mm,tip}];
		\chainin (es)    [join=by tip,join=with p5 by {vhpath, skip loop=-6mm,tip},join=with p2 by {skip loop=-6mm,tip}];
		\chainin (codes) [join=by tip];
  }
\end{tikzpicture}
};
\end{tikzpicture}
\caption{Data flow of Gabidulin decoding}
\label{fig:gflow}
\end{figure}
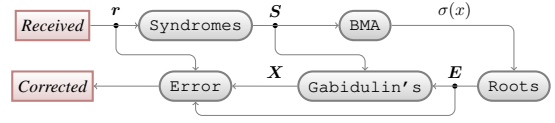

The decoding process of Gabidulin codes includes five major steps: syndrome computation, key equation solver, finding the root space, finding the error locators by Gabidulin's algorithm~\cite{Gabidulin85}, and finding error locations.
The data flow of Gabidulin decoding is shown in Figure~\ref{fig:gflow}.

Key equation solvers based on a modified Berlekamp--Massey algorithm (BMA)~\cite{Richter04} or a modified Welch--Berlekamp algorithm (WBA)~\cite{Loidreau05} have been proposed.
In this paper, we focus on the modified BMA due to its low complexity.

As in RS decoding, we can compute syndromes for Gabidulin codes as
$\bm{S}=(S_0,S_1,\dotsc,S_{d-2}) \triangleq \bm{Hr}$ for any received vector $\bm{r}$.
Then the syndrome polynomial $S(x)=\sum_{j=0}^{d-2}S_j x^{[j]}$ can be used to solve the key equation \cite[Theorem~3]{Richter04}
\begin{equation}
	\sigma(x)\otimes S(x) \equiv \omega(x) \bmod x^{[d-1]}
	\label{eqn:gkey}
\end{equation}
for the error span polynomial $\sigma(x)$, using the BMA.
Up to $t = \lfloor (d-1)/2\rfloor$ error values $E_j$'s can be obtained by finding a basis $E_0,E_1,\dotsc$ for the root space of $\sigma(x)$ using the methods in~\cite{Berlekamp68,Skachek08}.
Then we can find the error locators $X_j$'s corresponding to $E_j$'s by solving a system of equations
\begin{equation}
S_l = \sum_{j=0}^{\tau-1} X_j^{[l]} E_j, \quad l=0,1,\dotsc,d-2
\label{eqn:locator}
\end{equation}
where $\tau$ is the number of errors.
Gabidulin's algorithm~\cite{Gabidulin85} in Algorithm~\ref{alg:gabidulin} can be used to solve \eqref{eqn:locator}.
Finally, the error locations $\bm{L}_j$'s are obtained from $X_j$'s by solving
\begin{equation}
	X_j = \sum_{i=0}^{n-1} L_{j,i}h_i, \quad j = 0, 1, \dotsc, \tau-1.
	\label{eqn:location}
\end{equation}
\begin{algorithm}[Gabidulin's Algorithm~\cite{Gabidulin85}]
	\SetKwFunction{gabi}{}
	\gabi

	\KwIn{$S_0,S_1,\dotsc,S_{d-2}$ and $E_0,E_1,\dotsc,E_{\tau-1}$}
	\KwOut{$X_0,X_1,\dotsc,X_{\tau-1}$}
	\renewcommand{\labelenumi}{\ref{alg:gabidulin}.\theenumi}
	\begin{enumerate}
	\item Compute $\tau\times\tau$ matrices $\bm{A}$ and $\bm{Q}$ as
\begin{align*}
	A_{i,j}&= \begin{cases}
		E_j & i = 0\\
		0, & i \ne 0, j < i\\
		A_{i-1,j} - A_{i-1,i-1}(\frac{A_{i-1,j}}{A_{i-1,i-1}})^{[-1]} & i \ne 0, j \ge i
	\end{cases}\\
	Q_{i,j}&= \begin{cases}
		S_j & i = 0\\
		Q_{i-1,j} - A_{i-1,i-1}(\frac{Q_{i-1,j+1}}{A_{i-1,i-1}})^{[-1]} & \text{otherwise.}
	\end{cases}
\end{align*}
\item Compute $X_i$'s recursively as
	$X_{\tau-1} = Q_{\tau-1,0}/A_{\tau-1,\tau-1}$ and
	$X_{i} = (Q_{i,0}-\sum_{j=i+1}^{\tau-1}A_{i,j}X_j)/A_{i,i}$, for $i=\tau-2,\tau-3,\dots,0$.
\end{enumerate}
	\label{alg:gabidulin}
\end{algorithm}
In total, the decoding complexity of Gabidulin codes is roughly $O(n^2(1-R))$
operations over $\mathbb{F}_{q^m}$~\cite{Gadouleau08a}, where $R=k/n$ is the code rate, or $O(dm^3)$ operations over $\mathbb{F}_q$~\cite{Kschischang09}.
Note that all polynomials involved in the decoding process are linearized polynomials.

Gabidulin codes are often viewed as the counterpart in rank metric codes of the well-known RS codes. As shown in Table~\ref{tab:analogy}, an analogy between RS and Gabidulin codes can be established in many aspects. Such an analogy helps us understand the decoding of Gabidulin codes, and in some cases allows us to adapt innovations proposed for RS codes to Gabidulin codes.
\begin{table}[htbp]
	\centering

	\caption{Analogy between Reed--Solomon and Gabidulin Codes}
	\scalebox{0.9}{
	\begin{tabular}{|c|c|c|}
		\hline
		& Reed--Solomon & Gabidulin\\
		\hline
		Metric & Hamming & Rank\\
		\hline
		Ring of & Polynomials & Linearized Polynomials\\
		\hline
		Degree & $i$ & $[i]=q^i$\\
		\hline
		Key Operation & Polynomial Multiplication & Symbolic Product\\
		\hline
		Generation Matrix & $[g_j^i]$ & $[g_j^{[i]}]$\\
		\hline
		Parity Check Matrix & $[h_j^i]$ & $[h_j^{[i]}]$\\
		\hline
		Key Equation Solver & BMA & Modified BMA\\
		\hline
		Error Locations & Roots & Root Space Basis \\
		\hline
		Error Value Solver & Forney's Formula & Gabidulin's Algorithm\\
		\hline
	\end{tabular}
	}
	\label{tab:analogy}
\end{table}

\subsection{KK Codes  and Their Decoding}
By the \emph{lifting} operation~\cite{Silva08}, KK codes can be constructed from Gabidulin codes.
Lifting can also be seen as a generalization of the standard approach to random linear network coding~\cite{Ho06}, which transmits matrices in the form $\bm{X}=[\bm{I} \mid \bm{x}]$, where $\bm{X} \in \mathbb{F}_q^{n \times M}$, $\bm{x} \in \mathbb{F}_q^{n\times m}$, and $m = M - n$.

In practice, the packet length could be very long. To accommodate long packets based on the KK codes, very large $m$ and $n$ are needed, which results in prohibitively high complexity due to the huge field size of $\mathbb{F}_{q^m}$.
A low-complexity approach in~\cite{Silva08} suggested that instead of using a single long Gabidulin code, a Cartesian product of many short Gabidulin codes with the same distance can be used to construct constant-dimension codes for long packets via the lifting operation.

Let the received matrix be $\bm{Y} = [\bm{\hat{A}} \mid \bm{y} ]$, where $\bm{\hat{A}} \in \mathbb{F}_q^{N\times n}$ and $\bm{y}\in \mathbb{F}_q^{N\times m}$.
Note that we always assume the received matrix is full-rank~\cite{Silva08}.
The row and column rank deficiencies of $\bm{\hat{A}}$ are $\delta=N-\rank\bm{\hat{A}}$ and $\mu = n-\rank \bm{\hat{A}}$, respectively.
In the decoding algorithm of~\cite{Silva08}, the matrix $\bm{Y}$ is first turned into an RRE form, and then the RRE form of $\bm{Y}$ is expanded into
$\bar{\bm{Y}} = \bigl[\begin{smallmatrix} \bm{I}_{\mathcal{U}^c} & \bm{0}\\
	\bm{0} & \bm{I}_{\delta}
\end{smallmatrix}\bigr] \RRE(\bm{Y}) = \bigl[\begin{smallmatrix} \bm{I}_n+\bm{\hat{L}}\bm{I}_\mathcal{U}^T & \bm{r}\\
	\bm{0} & \bm{\hat{E}}
\end{smallmatrix}\bigr]$,
where $\mathcal{U}^c$ denotes the column positions of leading entries in the first $n$ rows of $\RRE(\bm{Y})$.
The tuple $(\bm{r}, \bm{\hat{L}}, \bm{\hat{E}})$ is called a \emph{reduction} of $\bm{Y}$~\cite{Silva08}.
It was proved~\cite{Silva08} that
	$d_S(\langle \bm{X}\rangle, \langle \bm{Y}\rangle) =
	 2\rank\bigl[\begin{smallmatrix}\bm{\hat{L}} & \bm{r}-\bm{x}\\
		\bm{0} & \bm{\hat{E}}
\end{smallmatrix}\bigr] - \mu - \delta$,
where $\mu = n - \rank \bm{\hat{L}}$ and $\delta = N - \rank \bm{\hat{L}}$. Now the decoding problem to minimize the subspace distance becomes a problem to minimize the rank distance.

For a KK code $\mathcal{C}$, the generalized rank decoding~\cite{Silva08} finds an error word $\bm{\hat{e}} = \argmin_{\bm{e} \in \bm{r} - \mathcal{C}} \rank\bigl[\begin{smallmatrix}\bm{\hat{L}} & \bm{e}\\
	0 & \bm{\hat{E}}
\end{smallmatrix}\bigr]$.
The error word $\bm{\hat{e}}$ is expanded as a summation of products of column and row vectors~\cite{Silva08} such that $\bm{\hat{e}} = \sum_{j=0}^{\tau-1} \bm{L}_j \bm{E}_j$.
Each term $\bm{L}_j \bm{E}_j$ is called either an \emph{erasure}, if $\bm{L}_j$ is known, or a \emph{deviation}, if $\bm{E}_j$ is known, or an \emph{error}, if neither $\bm{L}_j$ nor $\bm{E}_j$ is known.
In this general decoding problem, $\bm{L}$ has $\mu$ columns from $\bm{\hat{L}}$ and $\bm{E}$ has $\delta$ rows from $\bm{\hat{E}}$.
Given a Gabidulin code of minimum distance $d$, the corresponding KK code is able to correct $\epsilon$ errors, $\mu$ erasures, and $\delta$ deviations as long as if $2\epsilon+\mu+\delta < d$.

\begin{figure*}[htbp]
	\centering
\begin{tikzpicture}
\ifCLASSOPTIONonecolumn
	\node[scale=0.76]{
\else
	\node[scale=0.6]{
\fi
\begin{tikzpicture}[draw=black!50]
    \matrix[row sep=4mm, column sep=4mm] {
		& \node (th) [terminal] {$\times h$}; & \node (pe) [conn] {}; & \node (minpolye) [terminal] {MinPoly};\\
        \node (words) [nonterminal] {Received};&
		\node (rre) [terminal] {RRE};&
		\node (pl) [point] {}; & \node (minpolyl) [terminal] {MinPoly};&
        \node (p3) [conn] {}; & \node (spdu) [terminal] {SymProd};&
		\node (psdu) [point] {}; & \node (bma) [terminal] {BMA}; &
        \node (sigmaf) [conn] {}; & \node (spfd) [terminal] {SymProd}; &
		\node (psfd) [point] {};\\
		& & \node (pr) [conn] {}; & \node (synds) [terminal] {Syndromes};
		& \node (ps) [conn] {};\\
		\node (codes) [nonterminal] {Corrected};& \node (es) [terminal] {Error};& \node (x) [point] {}; & \node (gabidulin) [terminal] {Gabidulin's};& \node (e) [conn] {}; & \node (rs)   [terminal]    {Roots}; &
		\node (psigma) [point] {}; &
        \node (sps) [terminal] {SymProd};&
		\node (psigmau) [point] {}; &
        \node (minpolyb) [terminal] {MinPoly};&
		\node (pbeta) [point] {}; &
		\node (gabidulinfd) [terminal] {Gabidulin's};&
		\\
    };

	\node[above of=p3, xshift=-13pt,yshift=-20pt] (lambdau) {$\sigma_D(x)$};
	\node[above of=sigmaf, yshift=-20pt]{$\sigma_F(x)$};
	\node[above of=lambdau, yshift=2pt]{$\lambda_U(x)$};
	\node[above of=ps, yshift=-22pt]{$\bm{S}$};
	\node[above of=psdu, xshift=4pt,yshift=-20pt]{$S_{DU}(x)$};
	\node[above of=psfd, xshift=4pt,yshift=-20pt]{$S_{FD}(x)$};
	\node[above of=pbeta, yshift=-20pt]{$\bm{\beta}$};
	\node[above of=psigmau, yshift=-20pt]{$\sigma_U(x)$};
	\node[above of=psigma, yshift=-20pt]{$\sigma(x)$};
	\node[above of=pe, xshift=-5pt,yshift=-20pt]{$\bm{\hat{X}}$};
	\node[above of=rre, xshift=-5pt, yshift=-13pt]{$\bm{\hat{L}}$};
	\node[above of=e, yshift=-20pt]{$\bm{E}$};
	\node[above of=x, yshift=-20pt]{$\bm{X}$};
	\node[above of=pl, yshift=-20pt]{$\bm{\hat{E}}$};
	\node[above of=pr, yshift=-20pt]{$\bm{\hat{r}}$};
    { [start chain]
        \chainin (words);
        \chainin (rre) [join=by tip];
		{[start branch=hatl]
			\chainin (th) [join=by tip];
			\chainin (minpolye) [join=by tip];
		}
		{[start branch=synds]
        	\chainin (synds)[join=by {vhpath, tip}];
		}
		\chainin (minpolyl) [join=by tip];
		\chainin (spdu) [join=by tip,join=with synds by {hvpath,tip},join=with minpolye by {hvpath, tip}];
		\chainin (bma)   [join=by tip];
		\chainin (spfd) [join=by tip,join=with synds by {hvpath,tip},join=with p3 by {skip loop=6mm,tip}];
		\chainin (gabidulinfd);
		\draw[->] (spfd.0) to[hvpath, rounded corners] (gabidulinfd.50);
		\draw[->] (pe) to[hvpath, skip loop=6mm,rounded corners] (gabidulinfd.130);
		\chainin (minpolyb) [join=by tip];
		\chainin (sps) [join=by tip];
		\draw[->] (sigmaf) to[hvpath, skip loop=-6mm, rounded corners] (sps.50);
		\draw[->] (p3) to[hvpath, skip loop=-6mm, rounded corners] (sps.130);
		\chainin (rs)    [join=by tip];
		\chainin (gabidulin) [join=by tip,join=with ps by {hvpath, skip loop=-5mm, tip}];
		\chainin (es)    [join=by tip,join=with e by {skip loop=-6mm,tip},join=with pr by {skip loop=-5mm,tip}];
		\chainin (codes) [join=by tip];
  }
\end{tikzpicture}
};
\end{tikzpicture}
\caption{Data flow of KK decoding}
\label{fig:kflow}
\end{figure*}
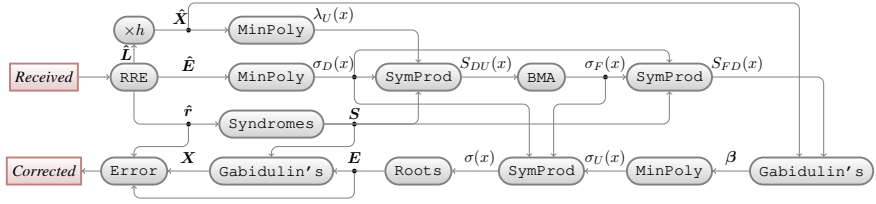

Algorithm~\ref{alg:general}  was proposed \cite{Silva08} for generalized decoding of the KK codes, and its data flow is shown in Figure~\ref{fig:kflow}. It requires $O(dm)$ operations in $\mathbb{F}_{q^m}$ \cite{Silva08}.

\begin{algorithm}[General Rank Decoding~\cite{Silva08}]
	\SetKwFunction{gGd}{}
	\gGd

	\KwIn{received tuple $(\bm{r}, \bm{\hat{L}}, \bm{\hat{E}})$}

	\KwOut{error word $\bm{\hat{e}}$}

	\renewcommand{\labelenumi}{\ref{alg:general}.\theenumi}
  \begin{enumerate}
  \item Compute $\bm{S} = \bm{Hr}$,
	  $\bm{\hat{X}} = \bm{\hat{L}}^T\bm{h}$,
	  $\lambda_U(x) = \minpoly(\bm{\hat{X}})$,
	  $\sigma_D(x) = \minpoly(\bm{\hat{E}})$, and $S_{DU}(x) = \sigma_D(x) \otimes S(x) \otimes {\zeta}_U(x)$, where ${\zeta}_U(x)$ is the $q$-reverse of $\lambda_U(x)$.
  \item Compute the error span polynomial:
  \begin{enumerate}
	  \item Use the modified BMA~\cite{Richter04} to solve the key equation $\sigma_F(x)\otimes S_{DU}(x) \equiv \omega(x) \bmod x^{[d-1]}$ such that $\deg \omega(x) < [\tau]$ where $\tau = \epsilon + \mu + \delta$.
    \item Compute $S_{FD}(x) = \sigma_{F}(x) \otimes \sigma_{D}(x) \otimes S(x)$.
	\item Use Gabidulin's algorithm~\cite{Gabidulin85} to find $\bm{\beta}$ that solves $S_{FD,l}=\sum_{j=0}^{\mu-1}X_j^{[l]}\beta_j, l = d-2, d-3, \dotsc, d-1-\mu$.
	\item Compute $\sigma_U(x) = \minpoly(\bm{\beta})$ followed by
    $\sigma(x) = \sigma_U(x) \otimes \sigma_{F}(x) \otimes \sigma_{D}(x)$.
  \end{enumerate}
  \item Find a basis $\bm{E}$ for the root space of $\sigma(x)$.
  \item Find the error locations:
  \begin{enumerate}
	  \item Solve $S_l = \sum_{j=0}^{\tau-1} X_j^{[l]} E_j, l = 0, 1,\dotsc, d-2$ using Gabidulin's algorithm~\cite{Gabidulin85} to find the error locators $X_0,X_1,\dotsc,X_{\tau-1} \in \mathbb{F}_{q^m}$.
	  \item Compute the error locations $\bm{L}_j$'s by solving \eqref{eqn:location}.
	  \item \label{item:error-word} Compute the error word $\bm{\hat{e}} = \sum_{j=1}^\tau \bm{L}_j \bm{E}_j$, where each $\bm{E}_j$ is the row expansion of $E_j$.
  \end{enumerate}
  \end{enumerate}
  \label{alg:general}
\end{algorithm}

\section{Computational Complexity Reduction}\label{sec:com}
In general, RLNC is carried out over $\mathbb{F}_{q}$, where $q$ is any prime power. That is, packets are treated as vectors over $\mathbb{F}_{q}$.
Since our investigation of computational complexities is for both software and hardware implementations of RLNC, where data are stored and transmitted in bits, we focus on RLNC over characteristic-$2$ fields in our work, i.e., $q$ is a power of two. In some cases, we further assume $q=2$,  as it leads to further complexity reductions.

\subsection{Finite Field Representation}
Finite field elements can be represented by vectors using different types of bases: polynomial basis, normal basis, and dual basis~\cite{Mastrovito91}. In rank metric decoders, most polynomials involved are linearized polynomials, and hence their evaluations and symbolic products require computing their $[i]$th powers. Suppose a field element is represented by a vector over $\mathbb{F}_{q}$ with respect to a normal basis, computing $[i]$th  powers ($i$ is a positive or negative integer) of the element is simply cyclic shifts of the corresponding vector by $i$ positions, which significantly reduces computational complexities. For example,  the computational complexity of Algorithm~\ref{alg:gabidulin} is primarily due to the following updates in Step~\ref{alg:gabidulin}.1:
\begin{equation}
\label{eqn:gabiq}
\begin{split}
	A_{i,j} &= A_{i-1,j}-(\frac{A_{i-1,j}}{A_{i-1,i-1}})^{[-1]}A_{i-1,i-1}\\
	Q_{i,j} &= Q_{i-1,j}-(\frac{Q_{i-1,j+1}}{A_{i-1,i-1}})^{[-1]}A_{i-1,i-1}
\end{split}
\end{equation}
which require divisions and computing $[-1]$th powers. With normal basis representation, $[-1]$th powers are obtained by a single cyclic shift. When $q=2$, they can be computed in an inversionless form
	$A_{i,j} = A_{i-1,j}-\bigl(A_{i-1,j}A_{i-1,i-1}\bigr)^{[-1]}$,
	$Q_{i,j} = Q_{i-1,j}-\bigl(Q_{i-1,j+1}A_{i-1,i-1}\bigr)^{[-1]}$,
which also avoids finite field divisions or inversions.
Thus using normal basis representation also reduces the complexity of Gabidulin's algorithm.

In addition to lower complexities of finite field arithmetic operations, normal basis representation leads to reduced complexities in the decoding of Gabidulin and KK codes for several reasons.
First, it was shown that using normal basis can facilitate the computation of symbolic product \cite{Gadouleau08a}.
Second, it was also suggested~\cite{Gadouleau08a} that solving \eqref{eqn:location} can be trivial using normal basis.
If $(h_0,h_1,\dotsc,h_{m-1})$ is a normal basis, the matrix $\bm{h}$, whose rows are vector representations of $h_i$'s with respect to the basis $h_i$'s, becomes an identity matrix with additional all-zero columns.
Hence solving \eqref{eqn:location} requires no computation.
These two complexity reductions were also observed in~\cite{Kschischang09}. Third, if a normal basis of $\mathbb{F}_{2^m}$ is used as $h_i$'s and $n=m$, the parity check matrix $\bm{H}$ in \eqref{eqn:check} becomes a cyclic matrix. Thus syndrome computation becomes part of a cyclic convolution of $(h_0,h_1,\dotsc,h_{m-1})$ and $\bm{r}$, for which fast algorithms are available~ (see, for example, \cite{Wagh83}). Using fast cyclic convolution algorithms are favorable when $m$ is large.

\subsection{Normal Basis Arithmetic Operations}
We also propose finite field arithmetic operations with reduced complexities, when normal basis representation is used.
When represented by vectors, the addition and subtraction of two elements are simply component-wise addition, which is straightforward to implement. For characteristic-$2$ fields $\mathbb{F}_{2^m}$, inverses can be obtained efficiently by a sequence of squaring and multiplying, since $\beta^{-1}=\beta^{2^m-2} = \beta^2 \beta^4 \dotso \beta^{2^{m-1}}$  for $\beta \in \mathbb{F}_{2^m}$ \cite{Mastrovito91}.
Since the $[i]$-th powers require no computation, the complexity of inversion in turn depends on that of multiplication.
Division can be implemented by a concatenation of inversion and multiplication: $\alpha/\beta=\alpha \cdot \beta^{-1}$, and hence the complexity of division also depends on that of multiplication in the end.

There are serial and parallel architectures for normal basis finite field multipliers.
To achieve high throughput in our decoder, we consider only parallel architectures.
Most normal basis multipliers are based on the Massey--Omura (MO) architecture \cite{Omura86,Mastrovito91}.
The complexity of a serial MO normal basis multiplier over $\mathbb{F}_{2^m}$, $C_N$, is defined as the number of terms $a_i b_j$ in computing a bit of the product $c=ab$, where $a=\sum_{i=0}^{m-1}a_i h_i \in \mathbb{F}_{2^m}$ and $b=\sum_{j=0}^{m-1}b_j h_j \in \mathbb{F}_{2^m}$ and $(h_0,h_1,\dotsc,h_{m-1})$ is a normal basis.
It has been shown \cite{Reyhani-Masoleh02} that a parallel MO multiplier over $\mathbb{F}_{2^m}$ needs $m^2$ AND gates and at most $m(C_N+m-2)/2$ XOR gates. For instance, for the fields $\mathbb{F}_{2^8}$ and $\mathbb{F}_{2^{16}}$, their $C_N$'s are minimized to 21 and 85, respectively~\cite{Mastrovito91}.
Using a common subexpression elimination algorithm \cite{Chen09}, we significantly reduce the number of XOR gates while maintaining the same critical path delays (CPDs) of one AND plus five XOR gates and one AND plus seven XOR gates as direct implementations, respectively.
Our results are compared to those in \cite{Mastrovito91,Reyhani-Masoleh02} in Table~\ref{tab:nb}, where we also provide the prime polynomial $P(x)$ for each field.
\begin{table}[htbp]
	\centering
	\caption{Complexities of bit-parallel normal basis multipliers over finite fields (For these two fields, all three implementations have the same CPD.)}
\ifCLASSOPTIONonecolumn
	\scalebox{1}{
\else
	\scalebox{0.7}{
\fi
	\begin{tabular}{|c|c|c|c|c|c|c|c|c|c|}
		\hline
		& \multirow{3}{*}{$P(x)$} & \multirow{2}{*}{AND} & \multicolumn{3}{c|}{XOR}\\
		\cline{4-6}
		& & & direct~\cite{Mastrovito91} & \cite{Reyhani-Masoleh02} & Ours\\
		\hline
		$\mathbb{F}_{2^m}$ & - & $m^2$ & $m(C_N-1)$ & $m(C_N+m-2)/2$ & -\\
		\hline
		$\mathbb{F}_{2^8}$ & $(\sum_{i=0}^8 x^i) - x^6 - x^4 - x^2$ & 64 & 160 & 108 & 88\\
		\hline
		$\mathbb{F}_{2^{16}}$ & $(\sum_{i=0}^{16} x^i) - x^{14} - x^9 - x^6 - x^4$ & 256 & 1344 & 792 & 491\\
		\hline
	\end{tabular}
	}
	
	\label{tab:nb}
\end{table}

The reduced gate count for normal basis multiplication is particularly important for hardware implementations of RLNC. This improvement is transparent to the complexity of decoders, in terms of finite field operations. When decoders for RLNC are realized in hardware, the reduced gate count for normal basis multiplication will be reflected in reduced area and power consumption.

\subsection{Inversionless BMA}\label{sec:bma}
The modified BMA for rank metric codes~\cite{Richter04} is similar to the BMA for RS codes except that polynomial multiplications are replaced by symbolic products. The modified BMA~\cite{Richter04} requires finite field divisions, which are more complex than other arithmetic operations.
Following the idea of inversionless RS decoder \cite{Burton71}, we propose an inversionless variant in Algorithm~\ref{alg:ibma}.
\begin{algorithm}

	\SetKwFunction{iBMA}{iBMA}

	\iBMA

	\KwIn{Syndromes $\bm{S}$}
	\KwOut{$\Lambda(x)$}
	
	\renewcommand{\labelenumi}{\ref{alg:ibma}.\theenumi}
	\begin{enumerate}
		\item Initialize: $\Lambda^{(0)}(x)=B^{(0)}(x)=x^{[0]}$, $\Gamma^{(0)}=1$, and $L=0$.
		\item For $r=0,1,\dotsc,2t-1$,
			\begin{enumerate}
				\item Compute the discrepancy $\Delta_r = \sum_{j=0}^L \Lambda_j^{(r)}S_{r-j}^{[j]}$.
				\item If $\Delta_r = 0$, then go to (e).
				\item Modify the connection polynomial: $\Lambda^{(r+1)}(x) = (\Gamma^{(r)})^{[1]}\Lambda^{(r)}(x) - \Delta_r x^{[1]}\otimes B^{(r)}(x)$.
				\item If $2L > r$, go to (e). Otherwise, $L=r+1-L$, $\Gamma^{(r+1)}=\Delta_r$, and $B^{(r)}(x) = \Lambda^{(r)}(x)$. Go to (a).
				\item Set $\Gamma^{(r+1)}=(\Gamma^{(r)})^{[1]}$ and $B^{(r+1)}(x)=x^{[1]}\otimes B^{(r)}(x)$.
			\end{enumerate}
		\item Set $\Lambda(x) = \Lambda^{(2t)}(x)$.
	\end{enumerate}
	\label{alg:ibma}
\end{algorithm}

Using a similar approach as in \cite{Burton71}, we prove that the output $\Lambda(x)$ of Algorithm~\ref{alg:ibma} is the same as $\sigma(x)$ produced by the modified BMA, except it is scaled by a constant $C=\prod_{i=0}^{t-1}(\Gamma^{(2i)})^{[1]}$.
However, this scaling is inconsequential since the two polynomials have the same root space.

Using normal basis, the modified BMA in \cite{Richter04} requires at most $\lfloor (d-2)/2\rfloor$ inversions, $(d-1)(d-2)$ multiplications, and $(d-1)(d-2)$ additions over $\mathbb{F}_{q^m}$~\cite{Gadouleau08a}.
Our inversionless version, Algorithm~\ref{alg:ibma}, requires at most $(3/2)d(d-1)$ multiplications and $(d-1)(d-2)$ additions.
Since a normal basis inversion is obtained by $m-1$ normal basis multiplications, the complexity of normal basis inversion is roughly $m-1$ times that of normal basis multiplication. Hence, Algorithm~\ref{alg:ibma} reduces the complexity considerably. Algorithm~\ref{alg:ibma} is also more suitable for hardware implementation, as shown in Section~\ref{sec:arch}.

\subsection{Finding the Root Space}
Instead of finding \emph{roots} of polynomials in RS decoding, we need to find the \emph{root spaces} of linearized polynomials in rank metric decoding. Hence the Chien search \cite{Berlekamp84} in RS decoding will have a high complexity for two reasons. First, it requires polynomial evaluations over the whole field, whose complexity is very high; Second, it cannot find a set of linearly independent roots.

A probabilistic algorithm to find the root space was proposed in~\cite{Skachek08}.
For Gabidulin codes, it can be further simplified as suggested in~\cite{Gadouleau08a}.
But hardware implementations of probabilistic algorithms require random number generators.
Furthermore, the algorithm in~\cite{Skachek08} requires symbolic long division, which is also not suitable for hardware implementations.
According to~\cite{Silva08}, the average complexity of the probabilistic algorithm in~\cite{Skachek08} is $O(dm)$ operations over $\mathbb{F}_{q^m}$, while that of Berlekamp's deterministic method~\cite{Berlekamp68} is $O(dm)$ operations in $\mathbb{F}_{q^m}$ plus $O(m^3)$ operations in $\mathbb{F}_q$.
Since their complexity difference is small, we focus on the deterministic method, which is much easier to implement.

Suppose we need to find the root space of a linearized polynomial $r(x)$, Berlekamp's deterministic method first evaluates the polynomial $r(x)$ on a basis of the field $(\alpha_0, \alpha_1, \dotsc, \alpha_{m-1})$ such that $v_i = r(\alpha_i), i = 0,1,\dotsc,m-1$.
Then it expands $v_i$'s in the base field as columns of an $m\times m$ matrix $\bm{V}$ and finds linearly independent roots $\bm{z}$ such that $\bm{Vz}=\bm{0}$.
Using the representation based on $(\alpha_0, \alpha_1, \dotsc, \alpha_{m-1})$, the roots $\bm{z}$ are also the roots of the given polynomial.
Finding $\bm{z}$ is to obtain the linear dependent combinations of the columns of $\bm{V}$, which can be done by Gaussian elimination.
\subsection{$n$-RRE Form}\label{sec:rre}
Given a received subspace spanned by a set of received packets, the input of Algorithm~\ref{alg:general} is a three-tuple, called a reduction of the received space represented by its generator matrix $\bm{Y}$; the three-tuple is obtained based on $\bm{Y}$ when it is in its RRE form \cite{Silva08}. Thus, before the decoding starts, preprocessing is performed on the received packets so as to obtain the RRE form of $\bm{Y}$.  We show that $\bm{Y}$ needs to satisfy only a relaxed constraint, which does not affect the decoding outcome, while leading to two advantages. First, the relaxed constraint results in reduced complexities in the preprocessing step. Second and more importantly, the relaxed constraint enables parallel processing of decoding KK codes based on Cartesian products.

We first define an $n$-RRE form for received matrices.
Given a matrix $\bm{Y} = [\bm{\hat{A}}\mid \bm{y}]$, where $\bm{\hat{A}} \in \mathbb{F}_q^{N\times n}$ and $\bm{y}\in \mathbb{F}_q^{N\times m}$, the matrix $\bm{Y}$ is in its $n$-RRE form as long as $\bm{\hat{A}}$ (its leftmost $n$ columns) is in its RRE form. Compared with the RRE form, the $n$-RRE form is more relaxed as it puts no constraints on the right part.
We note that an $n$-RRE form of a matrix is not unique.

We now show that the relaxed constraint does not affect the decoding.
Similar to \cite[Proposition~7]{Silva08}, we first show that a reduction based on $n$-RRE form of $\bm{Y}$ always exists.
Given $\bm{Y} = [\bm{\hat{A}}\mid \bm{y}]$ and $\RRE(\bm{\hat{A}})=\bm{R\hat{A}}$, where $\bm{R}$ represents the reducing row operations, the product $\bar{\bm{Y}}'=\bm{RY}= [\bm{B}' \mid \bm{Z}']$ is in its $n$-RRE form. We note that $\bm{B}' \in \mathbb{F}_q^{N\times n}$ and $\bm{Z} \in \mathbb{F}_q^{N\times m}$, where the column and row rank deficiency of $\bm{B}'$ are given by $\mu'=n - \rank \bm{B}'$ and $\delta'=N-\rank \bm{B}'$, respectively. We have the following result about the reduction based on  $\bar{\bm{Y}}'$.

\begin{lemma}
Let $\bar{\bm{Y}}'$ and $\mu'$ and $\delta'$ be defined as above. There exists a tuple $(\bm{r}', \bm{\hat{L}}', \bm{\hat{E}}') \in \mathbb{F}_q^{n\times m} \times \mathbb{F}_q^{n\times \mu'} \times \mathbb{F}_q^{\delta'\times m}$ and a set $\mathcal{U}'$ satisfying
$\abs{\mathcal{U}'}  = \mu'$,
$\bm{I}_{\mathcal{U}'}^T \bm{r}'  = 0$,
$\bm{I}_{\mathcal{U}'}^T \bm{\hat{L}}'  = -\bm{I}_{\mu' \times \mu'}$,
and $\rank \bm{\hat{E}}'  = \delta'$
so that $\Bigl \langle \bigl[\begin{smallmatrix} \bm{I}_n+\bm{\hat{L}}'\bm{I}_{\mathcal{U}'}^T & \bm{r}'\\
	0 & \bm{\hat{E}}'
\end{smallmatrix}\bigr] \Bigr \rangle= \langle \bar{\bm{Y}}' \rangle= \langle \bm{Y} \rangle$.
\label{lem:reduction}\end{lemma}
See Appendix~\ref{sec:proof1} for the proof of Lemma~\ref{lem:reduction}.  Lemma~\ref{lem:reduction} shows that we can find an alternative reduction based on $n$-RRE form of $\bm{Y}$, instead of an RRE form of $\bm{Y}$. The key of our alternative reduction of $\bm{Y}$ is that the reduction is mostly determined by the first $n$ columns of $\RRE(\bm{Y})$. Also, this alternative reduction does not come as a surprise. As shown in \cite[Proposition~8]{Silva08}, row operations on $\bm{\hat{E}}$ can produce alternative reductions. Next, we show that decoding based on our alternative reduction is the same as in \cite{Silva08}. Similar to \cite[Theorem~9]{Silva08}, we have the following results.

\begin{lemma}Let $(\bm{r}', \bm{\hat{L}}', \bm{\hat{E}}')$ be a reduction of $\bm{Y}$ determined by its $n$-RRE form, we have $d_S(\langle \bm{X}\rangle, \langle \bm{Y}\rangle) = 2\rank\bigl[\begin{smallmatrix}\bm{\hat{L}}' & \bm{r}'-\bm{x}\\
			\bm{0} & \bm{\hat{E}}'
		\end{smallmatrix}\bigr] - \mu' - \delta'.$\label{lem:equivalence}\end{lemma}

See Appendix~\ref{sec:proof2} for the proof.	Lemma~\ref{lem:equivalence} shows that the subspace decoding problem is equivalent to the generalized Gabidulin decoding problem with the alternative reduction $(\bm{r}', \bm{\hat{L}}', \bm{\hat{E}}')$, which is obtained from an $n$-RRE form of $\bm{Y}$.

Our alternative reduction leads to two advantages. First, it results in reduced complexity in preprocessing. Given a matrix $\bm{Y}$, the preprocessing needed to transform $\bm{Y}$ into its $n$-RRE form is only part of the preprocessing to transform $\bm{Y}$ into its RRE form. We can show that the maximal number of arithmetic operations in the former preprocessing is given by $(N-1)\sum_{i=0}^{\rank\bm{\hat{A}}-1}(n+m-i)$, whereas that of  the latter preprocessing is $(N-1)\sum_{i=0}^{\rank\left(\bm{Y}\right)-1}(n+m-i)$. Since $\rank \bm{Y} \geq \rank \bm{\hat{A}}$, the relaxed constraint leads to a lower complexity, and the reduction depends on $\rank \bm{Y}$ and $\rank \bm{\hat{A}}$.
Second, the reduction for $n$-RRE forms is completely determined by the $n$ leftmost columns of $\bm{Y}$ instead of the whole matrix, which greatly simplifies hardware implementations.
This advantage is particularly important for the decoding of constant-dimension codes that are lifted from Cartesian products of Gabidulin codes.
Since the row operations to obtain an $n$-RRE form depend on $\bm{\hat{A}}$ only, decoding $[\bm{\hat{A}} \mid \bm{y}_0 \mid \bm{y}_1 \mid \dotsb \mid \bm{y}_{l-1}]$ can be divided into parallel and smaller decoding problems whose inputs are $[\bm{\hat{A}} \mid \bm{y}_0], [\bm{\hat{A}} \mid \bm{y}_1], \dotsc, [\bm{\hat{A}} \mid \bm{y}_{l-1}]$. Thus, for these constant-dimension codes, we can decode in a serial manner with only one small decoder, or in a partly parallel fashion with more decoders, or even in a fully parallel fashion.
This flexibility allows tradeoffs between cost/area/power and throughput.
Furthermore, since the erasures $\bm{\hat{L}}$ are determined by $\bm{\hat{A}}$ and are the same for all $[\bm{A} \mid \bm{y}_i]$, the computation of $\bm{\hat{X}}$ and $\lambda_U(x)$
in Algorithm~\ref{alg:general} can be shared among these parallel decoding problems, thereby reducing overall complexity.

\subsection{Finding Minimal Linearized Polynomials}
Minimal linearized polynomials can be computed by solving systems of linear equations.
Given roots $\beta_0, \beta_1, \dotsc, \beta_{p-1}$, the minimal linearized polynomial $x^{[p]} + \sum_{i=0}^{p-1} a_i x^{[i]}$ satisfies
\begin{equation}
\begin{bmatrix}
	\beta_0^{[0]} & \beta_0^{[1]} & \dotsb & \beta_0^{[p-1]}\\
	\beta_1^{[0]} & \beta_1^{[1]} & \dotsb & \beta_1^{[p-1]}\\
	\vdots & \vdots & \ddots & \vdots\\
	\beta_{p-1}^{[0]} & \beta_{p-1}^{[1]} & \dotsb & \beta_{p-1}^{[p-1]}
\end{bmatrix}
\begin{bmatrix}
	a_0\\ a_1\\ \vdots\\ a_{p-1}
\end{bmatrix}
=
\begin{bmatrix}
	\beta_0^{[p]}\\ \beta_1^{[p]}\\ \vdots\\ \beta_{p-1}^{[p]}
\end{bmatrix}.
\label{eqn:minpoly}
\end{equation}
Thus it can be solved by Gaussian elimination over the extension field $\mathbb{F}_{q^m}$.
Gabidulin's algorithm is not applicable because the rows of the matrix are not the powers of the same element.

The complexity to solve \eqref{eqn:minpoly} is very high.
Instead, we reformulate the method from~\cite[Chap.~1, Theorem~7]{Ore33}.
The main idea of~\cite[Chap.~1, Theorem~7]{Ore33} is to recursively construct the minimal linearized polynomial using symbolic products instead of polynomial multiplications in polynomial interpolation.
Given linearly independent roots $w_0, w_1, \dotsc, w_{p-1}$, we can construct a series of linearized polynomials as:
$F^{(0)}(x) = x^{[0]}$ and $F^{(i+1)}(x) =  (x^{[1]}-(F^{(i)}(w_{i}))^{q-1}x^{[0]})\otimes F^{(i)}(x)$ for $i=0, 1, \cdots, p-1$.

Although the recursive method in \cite[Chap.~1, Theorem~7]{Ore33} is for $p$-polynomials, we can adapt it to linearized polynomials readily.
A serious drawback of \cite[Chap.~1, Theorem~7]{Ore33} is that the evaluation of $F^{(i)}(w_{i})$ has a rapidly increasing complexity when the degree of $F^{(i)}(x)$ gets higher.
To eliminate this drawback, we reformulate the algorithm so that the evaluation $F^{(i)}(w_{i})$ is done in a recursive way.
Our reformulated algorithm  is based on the fact that 
$F_{i}(w_{i+1}) = (x^{[1]}-(F_{i-1}(w_i))^{q-1}x^{[0]})\otimes F_{i-1}(w_{i+1})$.
Representing $F^{(i)}(w_{j})$ as $\gamma^{(i)}_j$, we obtain Algorithm~\ref{alg:minpoly}.

\begin{algorithm}[Minimal Linearized Polynomials]
	\SetKwFunction{MP}{}

	\MP

	\KwIn{Roots $w_0, w_1, \dotsc, w_{p-1}$}
	\KwOut{The minimal linearized polynomial $F^{(p)}(x)$}
	\renewcommand{\labelenumi}{\ref{alg:minpoly}.\theenumi}
	\begin{enumerate}
		\item Set $\gamma^{(0)}_j = w_j$, for $j= 0,1,\dotsc,p-1$ and $F^{(0)}(x)=x^{[0]}$.
		\item For $i=0, 1, \dotsc, p-1$,
			\begin{enumerate}
				\item If $\gamma^{(i)}_i = 0$, $F^{(i+1)}(x)= F^{(i)}(x)$ and $\gamma^{(i+1)}_j=\gamma^{(i)}_j$ for $j=i+1, i+2, \dotsc, p-1$; Otherwise, $F^{(i+1)}(x) = (F^{(i)}(x))^{[1]} - (\gamma^{(i)}_i)^{q-1}F^{(i)}(x)$ and $\gamma^{(i+1)}_j = (\gamma^{(i)}_j)^{[1]} - (\gamma^{(i)}_i)^{q-1} \gamma^{(i)}_j$ for $j=i+1, i+2, \dotsc, p-1$.
			\end{enumerate}
	\end{enumerate}
	\label{alg:minpoly}
\end{algorithm}
Since powers of $q$ require only cyclic shifting, the operations in Algorithm~\ref{alg:minpoly} are simple.
Also, Algorithm~\ref{alg:minpoly} does not require the roots to be linearly independent.
In Algorithm~\ref{alg:minpoly}, $F^{(i)}(w_j) = 0$ for $j= 0, 1, \dotsc, i-1$ and $\gamma^{(i+1)}_i = F^{(i)}(w_i)$.
If $w_0, w_1, \dotsc, w_j$ are linearly dependent, $\gamma^{(j)}_j=0$ and hence $w_j$ is ignored.
So Algorithm~\ref{alg:minpoly} integrates detection of linearly dependency at no extra computational cost.

Essentially, Algorithm~\ref{alg:minpoly} breaks down evaluations of high $q$-degree polynomials into evaluations of polynomials with $q$-degree of one. It avoids operations with very high complexity while maintaining  the same total complexity of the algorithm.

\section{Architecture Design}\label{sec:arch}
Aiming to reduce the storage requirement and total area as well as to improve the regularity of our decoder architectures, we further reformulate the steps in the decoding algorithms of both Gabidulin and KK codes. Again, we assume the decoder architectures are suitable for RLNC over $\mathbb{F}_{q}$, where $q$ is a power of two.

\subsection{High-Speed BMA Architecture}
To increase the throughput, regular BMA architectures with shorter CPD are necessary.
Following the approaches in~\cite{Sarwate01}, we develop two architectures based on Algorithm~\ref{alg:ibma}, which are analogous to the $\mathrm{riBM}$ and $\mathrm{RiBM}$ algorithms in \cite{Sarwate01}.

In Algorithm~\ref{alg:ibma}, the critical path is in step~3.2(a).
Note that $\Delta_r$ is the $r$th coefficient of the discrepancy polynomial $\Delta^{(r)}(x)=\Lambda^{(r)}(x)\otimes S(x)$.
By using $\Theta^{(r)}(x) = B^{(r)}(x)\otimes S(x)$, $\Delta^{(r+1)}(x)$ can be computed as
\begin{align}
	\Delta^{(r+1)}(x) &= \Lambda^{(r+1)}(x)\otimes S(x)\nonumber\\
		&= \bigl[(\Gamma^{(r)})^{[1]}\Lambda^{(r)}(x)-\Delta_r x^{[1]}\otimes B^{(r)}(x)\bigr]\otimes S(x)\nonumber\\
		&= (\Gamma^{(r)})^{[1]}\Delta^{(r)}(x)-\Delta_r x^{[1]}\otimes \Theta^{(r)}(x)\label{eqn:delta}
\end{align}
which has the same structure as step~3.2(c).
Hence this reformulation is more conducive to a regular implementation.

Given the similarities between step 3.2(a) and \eqref{eqn:delta}, $\Lambda(x)$ and $\Delta(x)$ can be combined together into one polynomial $\tilde\Delta(x)$.
Similarly, $B(x)$ and $\Theta(x)$ can be combined into one polynomial $\tilde\Theta(x)$.
These changes are incorporated in our RiBMA algorithm, shown in Algorithm~\ref{alg:Ribma}.
\begin{algorithm}

	\SetKwFunction{RiBMA}{RiBMA}

	\RiBMA

	\KwIn{Syndromes $\bm{S}$}
	\KwOut{$\Lambda(x)$}
	
	\BlankLine
	\renewcommand{\labelenumi}{\ref{alg:Ribma}.\theenumi}
	\begin{enumerate}
		\item Initialize: $\tilde\Delta^{(0)}(x) = \tilde\Theta^{(0)}(x) = \sum_{i=0}^{2t-1}S_i x^{[i]}$, $\Gamma^{(0)}=1$, $\tilde\Delta^{(0)}_{3t} = \tilde\Theta^{(0)}_{3t}=1$, and $b=0$.
		\item For $r = 0,1, \dotsc, 2t-1$,
			\begin{enumerate}
				\item Modify the combined polynomial: $\tilde\Delta^{(r+1)}(x) = \Gamma^{(r)}\tilde\Delta^{(r)}(x) - \tilde\Delta^{(r)}_0 \tilde\Theta^{(r)}(x)$;
				\item Set $b=b+1$;
				\item If $\tilde\Delta_0^{(r)}\ne0$ and $b > 0$, set $b=-b$, $\Gamma^{(r+1)} = \tilde\Delta^{(r)}_0$, and $\tilde\Theta^{(r)}(x) = \tilde\Delta^{(r)}(x)$;
				\item Set $\tilde\Delta^{(r+1)}(x) = \sum_{i=0}^{3t-1}\tilde\Delta^{(r+1)}_{i+1}x^{[i]}$, $\tilde\Theta^{(r)}(x) = \sum_{i=0}^{3t-1}\tilde\Theta^{(r)}_{i+1}x^{[i]}$;
				\item Set $\Gamma^{(r+1)}=(\Gamma^{(r)})^{[1]}$ and $\tilde\Theta^{(r+1)}(x) = x^{[1]} \otimes \tilde\Theta^{(r)}(x)$.
			\end{enumerate}
		\item Set $\Lambda(x) = \sum_{i=0}^t \tilde\Delta^{(2t)}_{i+t}x^{[i]}$.
	\end{enumerate}
	\label{alg:Ribma}
\end{algorithm}

\begin{figure}[htbp]
	\centering
	\begin{tikzpicture}
\ifCLASSOPTIONonecolumn
	\node[scale=0.9]{
\else
	\node[scale=0.6]{
\fi
\begin{tikzpicture}[
		hvpath/.style={to path={-| (\tikztotarget)}},
        vhpath/.style={to path={|- (\tikztotarget)}},
  		skip loop/.style={to path={- ++(0,#1) -| (\tikztotarget)}}]
	\matrix[row sep=10pt, column sep=20pt] {
	\node[pe] (pe0) {$\BE_0$}; & \node {$\dotsb$}; &
	\node[pe] (pe3) {$\BE_{t}$}; &
	\node {$\dotsb$}; &
	\node[pe] (pe02) {$\BE_{2t}$}; &
	\node {$\dotsb$}; &
	\node[pe] (pe7) {$\BE_{3t}$};
	& \node[pe] (ctrl) {BCtrl};\\
	};

	\draw[->] (ctrl.160) to[vhpath] (pe7.20);
	\draw[->] (ctrl.180) to[vhpath] (pe7.0);
	\draw[->] (ctrl.200) to[vhpath] (pe7.-20);
	\draw[->] (pe7.140) -- ++(-20pt,0);
	\draw[->] (pe7.160) -- ++(-20pt,0);
	\draw[->] (pe7.180) -- ++(-20pt,0);
	\draw[->] (pe7.200) -- ++(-20pt,0);
	\draw[->] (pe7.220) -- ++(-20pt,0);
	\draw[->] (pe3.140) -- ++(-20pt,0);
	\draw[->] (pe3.160) -- ++(-20pt,0);
	\draw[->] (pe3.180) -- ++(-20pt,0);
	\draw[->] (pe3.200) -- ++(-20pt,0);
	\draw[->] (pe3.220) -- ++(-20pt,0);
	\draw[->] (pe02.140) -- ++(-20pt,0);
	\draw[->] (pe02.160) -- ++(-20pt,0);
	\draw[->] (pe02.180) -- ++(-20pt,0);
	\draw[->] (pe02.200) -- ++(-20pt,0);
	\draw[->] (pe02.220) -- ++(-20pt,0);

	\draw[->] (pe0.160) -- ++(-20pt,0) -- ++(0, 50pt) to[hvpath] (ctrl.110);
	\draw[->] (pe0.180) -- ++(-15pt,0) -- ++(0, 50pt) to[hvpath] (ctrl.90);
	\draw[->] (pe0.200) -- ++(-10pt,0) -- ++(0, 50pt) to[hvpath] (ctrl.70);

	\draw[->] (pe3.120) -- ++(0, 10pt);
	\draw (pe3.120) ++(0, 15pt) node{$\Lambda_{0}$};

	\draw[->] (pe02.120) -- ++(0, 10pt);
	\draw (pe02.120) ++(0, 15pt) node{$\Lambda_{t}$};

	\draw[<-] (pe7.40) -- ++(10pt, 0);
	\draw (pe7.40) ++(15pt, 0) node{$0$};
	\draw[<-] (pe7.-40) -- ++(10pt, 0);
	\draw (pe7.-40) ++(15pt, 0) node{$0$};

	\draw[<-] (pe02.40) -- ++(20pt,0);
	\draw[<-] (pe02.20) -- ++(20pt,0);
	\draw[<-] (pe02.0) -- ++(20pt,0);
	\draw[<-] (pe02.-20) -- ++(20pt,0);
	\draw[<-] (pe02.-40) -- ++(20pt,0);
	\draw[<-] (pe3.40) -- ++(20pt,0);
	\draw[<-] (pe3.20) -- ++(20pt,0);
	\draw[<-] (pe3.0) -- ++(20pt,0);
	\draw[<-] (pe3.-20) -- ++(20pt,0);
	\draw[<-] (pe3.-40) -- ++(20pt,0);
	\draw[<-] (pe0.40) -- ++(20pt,0);
	\draw[<-] (pe0.20) -- ++(20pt,0);
	\draw[<-] (pe0.0) -- ++(20pt,0);
	\draw[<-] (pe0.-20) -- ++(20pt,0);
	\draw[<-] (pe0.-40) -- ++(20pt,0);
\end{tikzpicture}
};
\end{tikzpicture}
\caption{The RiBMA architecture}
\label{fig:Ribma}
\end{figure}
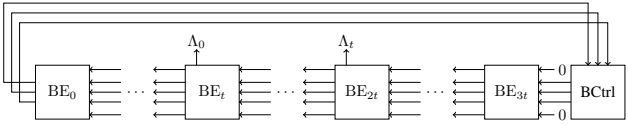

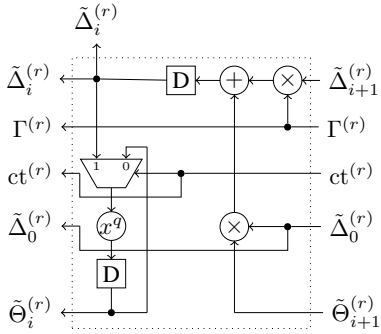
\begin{figure}[htbp]
	\centering
	\begin{tikzpicture}
\ifCLASSOPTIONonecolumn
	\node[scale=0.9]{
\else
	\node[scale=0.9]{
\fi
\begin{tikzpicture}[node distance=20pt]
	\matrix[draw, dotted, row sep=10pt, column sep=10pt] {
	& \node[reg] (D0) {D}; & \node[op] (add) {$+$}; & \node[op] (mul0) {$\times$};\\
	\node (dummy8) {}; & & & \node[conn] (conn9) {};\\
	\node[mux] (mux) {}; & \node[conn] (conn1) {}; & & \node (dummy1) {};\\
	& & \node[op] (mul1) {$\times$}; & \node[conn] (conn2) {};\\
	\node (dummy4) {}; & & & \node (dummy5) {};\\
	\node[conn] (conn3) {}; & & & \node (dummy6) {};\\
	};
	\node[right of=mul0, xshift=8pt] (deltar) {$\tilde{\Delta}_{i+1}^{(r)}$};
	\node[left of=dummy8, xshift=-13pt] (gammal) {$\Gamma^{(r)}$};
	\node[right of=conn9, xshift=5pt] (gammar) {$\Gamma^{(r)}$};
	\draw (mux.north west) node (dummy0) {};
	\draw (mux.south) node (dummy2) {};
	\draw (mux.north west) ++(0,-3pt) node {\tiny$1$};
	\draw (mux.north east) ++(0,-3pt) node {\tiny$0$};
	\node[op, below of=dummy2, yshift=4pt] (sq) {$x^{q}$};
	\node[reg, below of=sq] (D1) {D};
	\node[left of=conn3, xshift=-14pt] (thetal) {$\tilde{\Theta}_i^{(r)}$};
	\node[right of=dummy6, xshift=9pt] (thetar) {$\tilde{\Theta}_{i+1}^{(r)}$};

	\node[left of=mux, xshift=-14pt] (ctl) {$\ct^{(r)}$};
	\node[right of=dummy1, xshift=7pt] (ctr) {$\ct^{(r)}$};
	\node[left of=sq, xshift=-14pt] (deltall) {$\tilde{\Delta}_0^{(r)}$};
	\node[right of=conn2, xshift=7pt] (deltarr) {$\tilde{\Delta}_0^{(r)}$};
	\node[conn, above of=dummy0, yshift=13.5pt] (conn0) {};
	\node[left of=conn0, xshift=-8.5pt] (deltal) {$\tilde{\Delta}_i^{(r)}$};
	\node[above of=conn0, yshift=5pt] (deltat) {$\tilde{\Delta}_i^{(r)}$};

	\node[below of=deltar, yshift=-8pt] (dummy7) {};
	\node[right of=gammal, xshift=28pt] (dummy6) {};

	\draw[->] (conn0) -- (deltat);
	\draw[->] (conn0) -- (deltal);
	\draw[-] (D0) -- (conn0);
	\draw[->] (add) -- (D0);
	\draw[->] (mul0) -- (add);
	\draw[->] (deltar) -- (mul0);

	\draw[->] (conn0) -- (mux.north west);
	\draw[-] (conn3) -- (dummy6 |- conn3);
	\draw[-] (dummy6 |- conn3) -- (dummy7 -| dummy6);
	\draw[->] (dummy7 -| dummy6) to[hvpath] (mux.north east);
	\draw[->] (conn9) -- (mul0);

    \draw[-] (ctr) -- (conn1);
	\draw[->] (conn1) -- (mux.east);
	\draw[-] (conn1) to[hvpath, skip loop=-10pt] (mux.bottom right corner |- ctl);
	\draw[->] (mux.bottom right corner |- ctl) -- (ctl);

	\draw[-] (deltarr) -- (conn2);
	\draw[->] (conn2) -- (mul1);
	\draw[-] (conn2) to[hvpath, skip loop=-10pt] (mux.bottom right corner |- deltall);
	\draw[->] (mux.bottom right corner |- deltall) -- (deltall);

	\draw[->] (mul1) -- (add);

	\draw[->] (mux.south) -- (sq);
	\draw[->] (sq) -- (D1);
	\draw[-] (gammar) -- (conn9);
	\draw[->] (conn9) -- (gammal);
	\draw[->] (conn3) -- (thetal);
	\draw[-] (D1) -- (conn3);
    \draw[-] (thetar) -- (thetar -| mul1);
    \draw[->] (thetar -| mul1) -- (mul1);
\end{tikzpicture}
};
\end{tikzpicture}
\caption{The processing element $\BE_i$ ($x^{q}$ is a cyclic shift, and requires no hardware but wiring)}
\label{fig:pe}
\end{figure}

Following Algorithm~\ref{alg:Ribma}, we propose a systolic RiBMA architecture shown in Fig.~\ref{fig:Ribma}, which consists of $3t+1$ identical processing elements ($\BE$s), whose circuitry is shown in Fig.~\ref{fig:pe}.
The central control unit $\BCtrl$, the rightmost cell in Fig.~\ref{fig:Ribma}, updates $b$, generates the global control signals $\ct^{(r)}$ and $\Gamma^{(r)}$, and passes along the coefficient $\Lambda_0^{(r)}$.
The control signal $\ct^{(r)}$ is set to 1 only if $\tilde{\Delta}_0^{(r)} \ne 0$ and $k > 0$.
In each processing element, there are two critical paths, both of which consist of one multiplier and one adder over $\mathbb{F}_{2^m}$.

\subsection{Generalized BMA}
The key equation of KK decoding is essentially the same as \eqref{eqn:gkey}, but $\omega(x)$ has a $q$-degree less than $\tau$ instead of $\lfloor (d-1)/2 \rfloor$.
Actually, in KK decoding, we do not know the exact value of $\tau$ before solving the key equation.
All we need is to determine the maximum number of correctable errors $t'$ given $\mu$ erasures and $\delta$ deviations, which is given by $t'= \lfloor(d-1-\mu-\delta)/2\rfloor$.
Hence we adapt our BMA in Section~\ref{sec:bma} to KK decoding, as in Algorithm~\ref{alg:gRibma}.
To apply Algorithm~\ref{alg:gRibma} to Gabidulin decoding, we can simply use $\theta=\mu+\delta=0$.
\begin{algorithm}[Generalized RiBMA]

	\SetKwFunction{gRiBMA}{}

	\gRiBMA

	\KwIn{$\bm{S}$ and $\theta$}
	\KwOut{$\Lambda(x)$}
	
	\BlankLine
	\renewcommand{\labelenumi}{\ref{alg:gRibma}.\theenumi}
	\begin{enumerate}
		\item Initialize as follows: $t'=\lfloor(d-1-\theta)/2\rfloor$, $\tilde\Delta^{(0)}(x) = \tilde\Theta^{(0)}(x) = \sum_{i=\theta}^{\theta+2t'-1}S_i x^{[i]}$, $\tilde\Delta^{(0)}_{2t'+t} = \tilde\Theta^{(0)}_{2t'+t}=1$, $\Gamma^{(0)}=1$, and $b=0$.
		\item For $r = 0,1, \dotsc, 2t'-1$,
			\begin{enumerate}
				\item Modify the combined polynomial: $\tilde\Delta^{(r+1)}(x) = \Gamma^{(r)}\tilde\Delta^{(r)}(x) - \tilde\Delta^{(r)}_0 \tilde\Theta^{(r)}(x)$;
				\item Set $b=b+1$;
				\item If $\tilde\Delta_0^{(r)}\ne0$ and $b > 0$, set $b=-b$, $\Gamma^{(r+1)} = \tilde\Delta^{(r)}_0$, and $\tilde\Theta^{(r)}(x) = \tilde\Delta^{(r)}(x)$;
				\item Set $\tilde\Delta^{(r+1)}(x) = \sum_{i=0}^{2t'+t-1}\tilde\Delta^{(r+1)}_{i+1}x^{[i]}$, $\tilde\Theta^{(r)}(x) = \sum_{i=0}^{2t'+t-1}\tilde\Theta^{(r)}_{i+1}x^{[i]}$;
				\item Set $\Gamma^{(r+1)}=(\Gamma^{(r)})^{[1]}$ and $\tilde\Theta^{(r+1)}(x) = x^{[1]} \otimes \tilde\Theta^{(r)}(x)$.
			\end{enumerate}
		\item Set $\Lambda(x) = \sum_{i=0}^{t'} \tilde\Delta^{(2t')}_{i+t}x^{[i]}$.
	\end{enumerate}
	\label{alg:gRibma}
\end{algorithm}

Compared with Algorithm~\ref{alg:Ribma}, we replace $t$ by $t'$.
The variable $t'$ makes it difficult to design regular architectures.
By carefully initializing $\tilde\Delta^{(0)}(x)$ and $\tilde\Theta^{(0)}(x)$, we ensure that the desired output $\Lambda(x)$ is always at a fixed position of $\tilde{\Delta}^{(2t')}(x)$, regardless of $\mu+\delta$.
Hence, the only irregular part is the initialization.
The initialization of Algorithm~\ref{alg:gRibma} can be done by shifting in at most $\theta$ cycles.
Hence the RiBMA architecture in Fig.~\ref{fig:Ribma} can be adapted to the KK decoder and keep the same worse-case latency of $2t$ cycles.

\subsection{Gaussian Elimination}
We need Gaussian elimination to obtain $n$-RRE forms as well as to find root spaces.
Furthermore, Gabidulin's algorithm in Algorithm~\ref{alg:gabidulin} is essentially a smart way of Gaussian elimination, which takes advantage of the properties of the matrix.
The reduction (to obtain $n$-RRE forms) and finding the root space are Gaussian eliminations on matrices over $\mathbb{F}_q$, while Gabidulin's algorithm operates on matrices over $\mathbb{F}_{q^m}$.
In this section, we focus on Gaussian eliminations over $\mathbb{F}_q$ and Gabidulin's algorithm will be discussed in Section~\ref{sec:gabidulin}.

For high-throughput implementations, we adapt the pivoting architecture in \cite{Bogdanov06}, which
was developed for non-singular matrices over $\mathbb{F}_2$.
It always keeps the pivot element on the top-left location of the matrix, by cyclically shifting the rows and columns.
Our Gaussian elimination algorithm, shown in Algorithm~\ref{alg:ge}, has three key differences from  the pivoting architecture in \cite{Bogdanov06}. First, Algorithm~\ref{alg:ge} is applicable to matrices over any field. Second and more importantly, Algorithm~\ref{alg:ge} can be used for singular matrices.  This feature is necessary since singular matrices occur in the reduction for the RRE form and finding the root space. Third, Algorithm~\ref{alg:ge} is also flexible about matrix sizes, which are determined by the variable numbers of errors, erasures, and deviations.

\begin{algorithm}[Gaussian Elimination for Root Space]
	\SetKwFunction{GEf}{}
	\GEf

	\KwIn{$\bm{M} \in \mathbb{F}_{q}^{m\times m}$, whose rows are evaluations of $\sigma(x)$ over the normal basis, and $\bm{B}=\bm{I_m}$}
	\KwOut{Linearly independent roots of $\sigma(x)$}
	\BlankLine
	\renewcommand{\labelenumi}{\ref{alg:ge}.\theenumi}
	\begin{enumerate}
		\item Set $i=0$.
		\item For $j=0,1,\dotsc,m-1$
			\begin{enumerate}
				\item $l=1$
                \item While $M_{0,0} = 0$ and $l < m-i$\\
                    \hspace{3mm}$l=l+1$, $\shiftup(\bm{M}, i)$, and $\shiftup(\bm{B}, i)$.
				\item If $M_{0,0}$ is not zero, $\eliminate(\bm{M})$, $\reduce(\bm{B}, \bm{M})$, and $i=i+1$; Otherwise, $\shiftleft(\bm{M})$.
			\end{enumerate}
		\item The first $m-i$ rows of $\bm{M}$ are all zeros and the first $m-i$ rows of $\bm{B}$ are roots.
	\end{enumerate}
	\label{alg:ge}
\end{algorithm}
The $\eliminate$ and $\shiftup$ operations are quite similar to those in~\cite[Algorithm~2]{Bogdanov06}.
In $\eliminate(\bm{M})$, for $0 \le j < m$, $M_{i,j} = M_{0,0}M_{i+1,(j+1) \bmod m} - M_{i+1,0}M_{0,(j+1)\bmod m}$ for $0 \le i < m-1$, and $M_{m-1,j} = M_{0, (j+1) \bmod m}$. Note that a cyclic row shift and a cyclic column shift are already embedded in the $\eliminate$ operation. In the $\shiftup(\bm{M}, \rho)$ operation, the first row is moved to the $(m-1-\rho)$th row while the second to the $(m-1-\rho)$th rows are moved up.
That is, for $0 \le j < m$, $M_{i,j} = M_{0,j}$ if $i = m-1-\rho$, and $M_{i,j} = M_{i+1, j}$ for $0 \leq i \leq m-2-\rho$. The operation $\reduce(\bm{B}, \bm{M})$ essentially mimics all row operations in $\eliminate$ without the column shift: for $0 \le j < m$, $B_{i,j} = M_{0,0}B_{i+1,j} - M_{i+1,0}B_{0,j}$ for $0 \le i < m-1$, and $B_{m-1,j} = B_{0, j}$.
In the $\shiftleft$ operation, all columns are cyclicly shifted to the left.
In other words, for all $0 \le i < m$ and $0 \le j< m$, $M_{i,j} = M_{i, (j+1)\bmod m}$.
By adding a $\shiftleft$ operation, Algorithm~\ref{alg:ge} handles both singular and non-singular matrices while~\cite[Algorithm~2]{Bogdanov06} works for non-singular matrices only. Since $\bm{B}$ is always full rank, the roots obtained are guaranteed to be linearly independent.

We can get the root space using Algorithm~\ref{alg:ge}, and we can also use it in KK decoding to reduce the received vector to an $n$-RRE form.
However, Algorithm~\ref{alg:ge} only produces $\bm{\hat{E}}'$. We extend it to Algorithm~\ref{alg:ege} below so as to obtain $\bm{\hat{L}}'$ simultaneously.

\begin{algorithm}[Gaussian Elimination for $n$-RRE Forms]
	\SetKwFunction{eGE}{}
	\eGE

	\KwIn{$N\times n$ matrix $\bm{\hat{A}}$ and $N \times m$ matrix  $\bm{y}$}
	\KwOut{$\bm{\hat{L}}'$, $\bm{\hat{E}}'$, $\bm{r}'$, and $\mu'$}
	\BlankLine
	\renewcommand{\labelenumi}{\ref{alg:ege}.\theenumi}
	\begin{enumerate}
		\item Set $i=0$, $\mathcal{U}'$ and $\bm{\hat{L}}$ as empty.
		\item For each column $j=0,1,\dotsc,n-1$
			\begin{enumerate}
				\item $l=1$
				\item While $\hat{A}_{0,0} = 0$ and $l < n-i$

					$l=l+1$, $\shiftup(\bm{\hat{A}}, i)$, $\shiftup(\bm{y}, i)$, $\shiftup(\bm{\hat{L}}', i)$.
				\item If $\hat{A}_{0,0}$ is not zero, $\eliminate(\bm{\hat{A}})$, $\reduce(\bm{y}, \bm{\hat{A}})$, $\shiftup(\bm{\hat{L}}', 0)$, $i=i+1$.
				\item Otherwise, $\shiftleft(\bm{\hat{A}})$, append the first column of $\bm{\hat{A}}$ to $\bm{\hat{L}}'$, set the top-right element of $\bm{\hat{L}}'$ to one, and add $j$ to $\mathcal{U}'$.
			\end{enumerate}
		\item Set $\mu' = n-i$. The deviations $\bm{\hat{E}}'$ are given by the first $\mu'$ rows of $\bm{y}$.
		\item For each column $j \in \mathcal{U}'$, $\shiftup(\bm{\hat{L}}', j)$ and $\shiftup(\bm{y}, j)$.
		\item The received vector $\bm{r}'$ is given by $\bm{y}$.
	\end{enumerate}
	\label{alg:ege}
\end{algorithm}

In Algorithm~\ref{alg:ege}, we incorporate the extraction of $\bm{\hat{L}}'$, $\bm{\hat{E}}'$, and $\bm{r}'$ into Gaussian elimination.
Our architecture has the same worst-case latency as Algorithm~\ref{alg:ge} and requires no extra cycles to extract $\bm{\hat{L}}$ out of the $n$-RRE form. Hence the throughput also remains the same.

Algorithm~\ref{alg:ge} is implemented by the regular architecture shown in Fig.~\ref{fig:gaussian}, which is a two-dimensional array of $m\times 2m$ processing elements ($\GE$'s). The leftmost $m$ columns of processing elements correspond to $\bm{M}$, and the rightmost $m$ columns $\bm{B}$.
Algorithm~\ref{alg:ege} can be implemented with the same architecture with $N\times (n+m)$ $\GE$'s. The leftmost $n$ columns of processing elements correspond to $\bm{\hat{A}}$, and the rightmost $m$ columns $\bm{y}$.
The elements for $\bm{\hat{L}}'$ are omitted in the figure.
The circuitry of the processing element $\GE$ is shown in Fig.~\ref{fig:ge}.
The control signal $\ct_i$ for row $i$ chooses from five inputs based on the operation: keeping the value, $\shiftleft$, $\eliminate$ (or $\reduce$), and $\shiftup$ (using the first row or the next row).
\begin{figure*}[htbp]
	\centering
	\begin{tikzpicture}
		\node[scale=0.8]{
	\begin{tikzpicture}[font=\fontsize{6pt}{6pt}]
		\matrix[row sep=15pt, column sep=10pt] {
		\node[peg] (ctrl) {GCtrl}; & \node[peg] (ge00) {$\GE_{0,0}$}; & \node[peg] (ge01) {$\GE_{0,1}$}; & \node{$\dotsb$}; & \node[peg] (ge0n) {$\GE_{0,m-1}$}; & \node[peg] (ge0n0) {$\GE_{0,m}$}; & \node[peg] (ge0n1) {$\GE_{0,m+1}$}; & \node{$\dotsb$}; & \node[peg] (ge0nn) {$\GE_{0,2m-1}$}; \\
		\node{$\dots$}; & \node[peg] (ge10) {$\GE_{1,0}$}; & \node[peg] (ge11) {$\GE_{1,1}$}; & \node{$\dotsb$}; & \node[peg] (ge1n) {$\GE_{1,m-1}$}; & \node[peg] (ge1n0) {$\GE_{1,m}$}; & \node[peg] (ge1n1) {$\GE_{1,m+1}$}; & \node{$\dotsb$}; & \node[peg] (ge1nn) {$\GE_{1,2m-1}$};\\
		\node{$\ddots$}; & \node{$\vdots$}; & \node{$\vdots$}; & \node{$\ddots$}; & \node{$\vdots$}; & \node{$\vdots$}; & \node{$\ddots$}; & \node{$\ddots$}; & \node {$\vdots$};\\\\
		& \node[peg] (gen0) {$\GE_{m-1,0}$}; & \node[peg] (gen1) {$\GE_{m-1,1}$}; & \node{$\dotsb$}; & \node[peg] (genn) {$\GE_{m-1,m-1}$}; & \node[peg] (genn0) {$\GE_{m-1,m}$}; & \node[peg] (genn1) {$\GE_{m-1,m+1}$}; & \node{$\dotsb$}; & \node[peg] (gennn) {$\GE_{m-1,2m-1}$}; \\
		};
		\draw[-] (ge00.70) -- ++(0, 5pt) node[conn] (g00) {};
		\draw[->] (g00) -- ++(18pt,0) to[vhpath] (ge01.150);
		\draw[->] (ge00.-30) -- (ge01.-150);
		\draw[<-] (ge00.0) -- (ge01.-180);
		\draw(ge10.70) ++(0, 5pt) node[conn] (10c) {};
		\draw[->] (10c) -- ++(18pt,0) to[vhpath] (ge11.150);
		\draw(ge10.-70) ++(0, -10pt) node[conn] (10c1) {};
		\draw[->] (10c1) -- ++(18pt,0) -- ++(0,-13pt) -- ++(5pt,0);
		\draw[->] (ge10.-30) -- (ge11.-150);
		\draw[<-] (ge10.0) -- (ge11.-180);
		\draw[->] (ge11.135) -- (ge00.-45);
		\draw[<-] (ge00.-70) -- (ge10.70);
		\draw[<-] (ge01.-70) -- (ge11.70);
		\draw[<-] (ge0n.-70) -- (ge1n.70);

		\draw[->] (ge01.30) -- ++(10pt,0);
		\draw[->] (ge01.-30) -- ++(10pt,0);
		\draw[<-] (ge01.0) -- ++(10pt,0);
		\draw[<-] (ge0n.150) -- ++(-10pt,0);
		\draw[<-] (ge0n.210) -- ++(-10pt,0);
		\draw[->] (ge0n.180) -- ++(-10pt,0);

		\draw[->] (ge11.30) -- ++(10pt,0);
		\draw[->] (ge11.-30) -- ++(10pt,0);
		\draw[<-] (ge11.0) -- ++(10pt,0);
		\draw[<-] (ge1n.150) -- ++(-10pt,0);
		\draw[<-] (ge1n.210) -- ++(-10pt,0);
		\draw[->] (ge1n.180) -- ++(-10pt,0);
		\draw[->] (gen1.30) -- ++(10pt,0);
		\draw[->] (gen1.-30) -- ++(10pt,0);
		\draw[<-] (gen1.0) -- ++(10pt,0);
		\draw[<-] (genn.150) -- ++(-10pt,0);
		\draw[<-] (genn.210) -- ++(-10pt,0);
		\draw[->] (genn.180) -- ++(-10pt,0);

		\draw(gen0.70) ++(0, 5pt) node[conn] (n0c) {};
		\draw[->] (n0c) -- ++(18pt,0) to[vhpath] (gen1.150);
		\draw[->] (gen0.-30) -- (gen1.-150);
		\draw[<-] (gen0.0) -- (gen1.-180);

		\draw[->] (gen0.70) -- ++(0,15pt);
		\draw[->] (gen1.70) -- ++(0,15pt);
		\draw[->] (genn.70) -- ++(0,15pt);

		\draw[<-] (ge10.-70) -- ++(0,-15pt);
		\draw[->] (ge10.-90) -- ++(0,-15pt);
		\draw[<-] (ge11.-70) -- ++(0,-15pt);
		\draw[->] (ge11.-90) -- ++(0,-15pt);
		\draw[<-] (ge1n.-70) -- ++(0,-15pt);
		\draw[->] (ge1n.-90) -- ++(0,-15pt);
		
		\draw[<-] (gen0.90) -- ++(0,15pt);
		\draw[<-] (gen1.90) -- ++(0,15pt);
		\draw[<-] (genn.90) -- ++(0,15pt);

		\draw[->] (ctrl.-30) -- (ge00.-150);
		\draw[->] (ctrl.-60) to[vhpath] (ge10.-150);
		\draw[->] (ctrl.-130) to[vhpath] (gen0.-150);

		\draw[->] (ge00.-90) -- (ge10.90);
		\draw[->] (ge01.-90) -- (ge11.90);
		\draw[->] (ge0n.-90) -- (ge1n.90);

		\draw[<-] (ge01.-45) -- ++(10pt,-15pt);
		\draw[->] (ge1n.135) -- ++(-10pt,15pt);
		\draw[<-] (ge10.-45) -- ++(10pt,-15pt);
		\draw[<-] (ge11.-45) -- ++(10pt,-15pt);
		\draw[->] (gen1.135) -- ++(-10pt,15pt);
		\draw[->] (genn.135) -- ++(-10pt,15pt);

		\draw[->] (g00) -- ++(-37pt,0pt) to[vhpath] (ge10.140);
		\node[conn,left of=ge10,xshift=0pt,yshift=19.5pt] (g10) {};
		\node[conn,below of=g10,yshift=-28pt] (g20) {};
		\draw[->] (g20) -- ++(5pt,0);
		\draw[->] (g10) to[vhpath] (gen0.140);
		\draw[->] (ge10.40) -- (ge11.140);
		\draw[->] (gen0.40) -- (gen1.140);
		\draw[->] (ge11.40) -- ++(10pt,0pt);
		\draw[->] (gen1.40) -- ++(10pt,0pt);
		\draw[<-] (ge1n.140) -- ++(-10pt,0pt);
		\draw[<-] (genn.140) -- ++(-10pt,0pt);

		\draw[->] (ge0n0.70) -- ++(0, 5pt) -- ++(18pt,0) to[vhpath] (ge0n1.150);
		\draw[->] (ge0n0.180) -- ++(-3pt, 0) -- ++(0, -32pt) -- ++(53pt, 0) to[vhpath] (ge0n0.0);
		\draw[->] (ge1n0.180) -- ++(-3pt, 0) -- ++(0, -32pt) -- ++(53pt, 0) to[vhpath] (ge1n0.0);
		\draw[->] (genn0.180) -- ++(-3pt, 0) -- ++(0, -32pt) -- ++(53pt, 0) to[vhpath] (genn0.0);
		\draw[->] (ge0n1.180) -- ++(-3pt, 0) -- ++(0, -32pt) -- ++(53pt, 0) to[vhpath] (ge0n1.0);
		\draw[->] (ge1n1.180) -- ++(-3pt, 0) -- ++(0, -32pt) -- ++(53pt, 0) to[vhpath] (ge1n1.0);
		\draw[->] (genn1.180) -- ++(-3pt, 0) -- ++(0, -32pt) -- ++(53pt, 0) to[vhpath] (genn1.0);
		\draw[->] (ge0n0.-90) -- (ge1n0.90);
		\draw[->] (ge0n1.-90) -- (ge1n1.90);
		\draw[<-] (ge0n0.-70) -- (ge1n0.70);
		\draw[<-] (ge0n1.-70) -- (ge1n1.70);
		\draw[->] (ge1n0.135) -- ++(-4pt,4pt) -- ++(55pt,0) -- (ge0n0.-45);
		\draw[->] (ge1n1.135) -- ++(-4pt,4pt) -- ++(55pt,0) -- (ge0n1.-45);
		\draw[->] (ge0n.-30) -- (ge0n0.-150);
		\draw[->] (ge0n0.-30) -- (ge0n1.-150);
		\draw[->] (ge1n.-30) -- (ge1n0.-150);
		\draw[->] (ge1n0.-30) -- (ge1n1.-150);
		\draw[->] (genn.-30) -- (genn0.-150);
		\draw[->] (genn0.-30) -- (genn1.-150);
		\draw[->] (ge1n.30) -- (ge1n0.150);
		\draw[->] (ge1n0.30) -- (ge1n1.150);
		\draw[->] (genn.30) -- (genn0.150);
		\draw[->] (genn0.30) -- (genn1.150);
		\draw[->] (ge1n.40) -- (ge1n0.140);
		\draw[->] (ge1n0.40) -- (ge1n1.140);
		\draw[->] (genn.40) -- (genn0.140);
		\draw[->] (genn0.40) -- (genn1.140);
		\draw[<-] (ge1n0.-70) -- ++(0,-15pt);
		\draw[->] (ge1n0.-90) -- ++(0,-15pt);
		\draw[<-] (ge1n1.-70) -- ++(0,-15pt);
		\draw[->] (ge1n1.-90) -- ++(0,-15pt);
		\draw[->] (genn0.70) -- ++(0,15pt);
		\draw[<-] (genn0.90) -- ++(0,15pt);
		\draw[->] (genn1.70) -- ++(0,15pt);
		\draw[<-] (genn1.90) -- ++(0,15pt);
		\draw[->] (genn0.135) -- ++(-4pt,4pt) -- ++(55pt,0) -- ++(-4pt,12pt);
		\draw[->] (genn1.135) -- ++(-4pt,4pt) -- ++(55pt,0) -- ++(-4pt,12pt);
		\draw[<-] (ge1n0.-45) -- ++(4pt,-12pt) -- ++(-55pt,0) -- ++(4pt,-4pt);
		\draw[<-] (ge1n1.-45) -- ++(4pt,-12pt) -- ++(-55pt,0) -- ++(4pt,-4pt);
		\draw[->] (ge0n1.30) -- ++(10pt,0);
		\draw[->] (ge1n1.30) -- ++(10pt,0);
		\draw[->] (genn1.30) -- ++(10pt,0);
		\draw[->] (ge1n1.40) -- ++(10pt,0);
		\draw[->] (genn1.40) -- ++(10pt,0);
		\draw[->] (ge0n1.-30) -- ++(10pt,0);
		\draw[->] (ge1n1.-30) -- ++(10pt,0);
		\draw[->] (genn1.-30) -- ++(10pt,0);
		\draw[<-] (ge0nn.150) -- ++(-10pt,0);
		\draw[<-] (ge0nn.210) -- ++(-10pt,0);
		\draw[<-] (ge1nn.140) -- ++(-10pt,0);
		\draw[<-] (ge1nn.150) -- ++(-10pt,0);
		\draw[<-] (ge1nn.210) -- ++(-10pt,0);
		\draw[<-] (gennn.140) -- ++(-10pt,0);
		\draw[<-] (gennn.150) -- ++(-10pt,0);
		\draw[<-] (gennn.210) -- ++(-10pt,0);
		\draw[->] (gennn.135) -- ++(-4pt,4pt) -- ++(55pt,0) -- ++(-4pt,12pt);
		\draw[<-] (ge1nn.-45) -- ++(4pt,-12pt) -- ++(-55pt,0) -- ++(4pt,-4pt);
		\draw[->] (ge1nn.135) -- ++(-4pt,4pt) -- ++(55pt,0) -- (ge0nn.-45);
		\draw[->] (ge0nn.180) -- ++(-3pt, 0) -- ++(0, -32pt) -- ++(53pt, 0) to[vhpath] (ge0nn.0);
		\draw[->] (ge1nn.180) -- ++(-3pt, 0) -- ++(0, -32pt) -- ++(53pt, 0) to[vhpath] (ge1nn.0);
		\draw[->] (gennn.180) -- ++(-3pt, 0) -- ++(0, -32pt) -- ++(53pt, 0) to[vhpath] (gennn.0);
		\draw[<-] (ge1nn.-70) -- ++(0,-15pt);
		\draw[->] (ge1nn.-90) -- ++(0,-15pt);
		\draw[->] (gennn.70) -- ++(0,15pt);
		\draw[<-] (gennn.90) -- ++(0,15pt);
		\draw[->] (ge0nn.-90) -- (ge1nn.90);
		\draw[<-] (ge0nn.-70) -- (ge1nn.70);
	\end{tikzpicture}
	};
	\end{tikzpicture}
	\caption{Regular architecture for Gaussian elimination}
	\label{fig:gaussian}
\end{figure*}
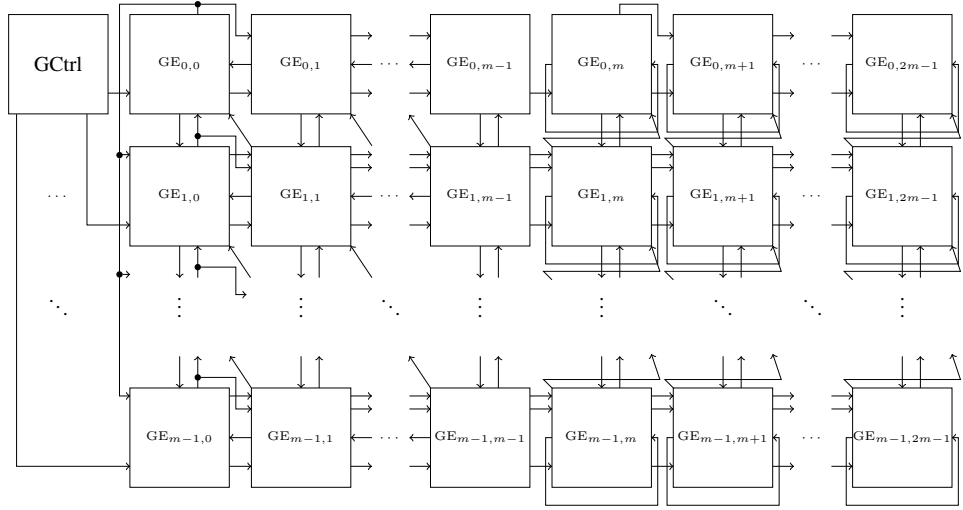
\begin{figure}[htbp]
	\centering
		\begin{tikzpicture}
\ifCLASSOPTIONonecolumn
	\node[scale=0.9]{
\else
	\node[scale=0.9]{
\fi
\begin{tikzpicture}[node distance=20pt]
	\matrix[draw, dotted, row sep=15pt, column sep=10pt] {
	\node[point] (do) {}; & & & & & \node[point] (uc) {};\\
	&\node[smux] (mux0) {};\\
	& & \node[reg] (D) {D}; & \node[emux] (mux1) {};\\
	& & & & \node[point] (bc) {}; & \node[point] (dc) {};\\
	};
	\draw (mux0.south west) ++(0, -15pt) node[op] (inv) {$+$};
	\node[op,below of=inv, yshift=-3pt] (mul0) {$\times$};
	\node[op,above of=D,xshift=5pt,yshift=10pt] (mul1) {$\times$};
	\node[right of=D,xshift=2pt,rotate=90] (txt) {\tiny{MUX}};
	\draw[->] (mul1) to[vhpath] (inv);
	\draw[->] (mul0) -- (inv);
	\node[left of=D, conn, xshift=-3pt] (r) {};
	\draw[->] (mux1.west) -- (D);
	\draw[-] (D) -- (r);
	\draw[->] (r) -- (mux0.south east);
	\draw[->] (inv) -- (mux0.south west);
	\node[conn, above of=r] (r0) {};
	\node[conn, right of=mul1,xshift=13.5pt] (r1) {};
	\draw[->] (r1) -- (mul1);
	\node[below of=mux1,conn] (cc) {};
	\node[below of=bc,yshift=10pt] (bi) {};
	\node[left of=bi, xshift=5pt] {$M_{i+1,j}$};
	\draw[->] (bc) to[vhpath] (mux1.bottom left corner);
	\draw[-] (bi) -- (bc);
	\node[right of=mux1, xshift=33pt] (ri) {$M_{i, j+1}$};
	\draw[->] (ri) -- (mux1);
	\draw[->] (dc) to[vhpath] (mux1.south east);
	\node[above of=uc, xshift=-5pt, yshift=-10pt] (ui) {};
	\node[right of=cc, xshift=25pt] (co) {$\ct_i$};
	\node[left of=cc, xshift=-72pt] (ci) {$\ct_i$};
	\draw[->] (ci) -- (co);
	\draw[->] (cc) -- (mux1);
	\node[coordinate,left of=r,xshift=3pt] (lc) {};
	\node[left of=r,xshift=-30pt] (lo) {$M_{i,j}$};
	\node[below of=dc, xshift=10pt, yshift=10pt] (di) {};
	\node[right of=di] {$M'_{i+1,j+1}$};
	\draw[-] (di) -- (dc);
	\draw[->] (lc) -- (lo);
	\node[conn, above of=bi, yshift=38pt] (uoc) {};
	\draw[-] (r0) -- (uoc);
	\draw[->] (uoc) to[vhpath] (mux1.bottom right corner);
	\node[above of=uoc, yshift=28pt] (uo) {};
	\draw[->] (uoc) -- (uo);
	\node[left of=uo, xshift=10pt] {$M_{i,j}$};
	\node[above of=do, xshift=-8pt, yshift=-10pt] (doo) {};
	\node[left of=doo, xshift=10pt] {$M'_{i,j}$};
	\draw[-] (mux0.north) to[vhpath] (do);
	\draw[->] (do) -- (doo);
	\node[conn,left of=mux0] (lcc) {};
	\draw[-] (r) -- (lc);
	\node[left of=lcc, xshift=-3pt] (lc) {$M_{i,0}$};
	\node[above of=lc,yshift=-4pt] (p0) {$M_{0,0}$};
	\draw[->] (lc) -- (mux0.west);
	\node[right of=mux0, xshift=80pt] (lco) {$M_{i,0}$};
	\node[above of=lco,yshift=-4pt] (p1) {$M_{0,0}$};
	\draw[->] (p0) -- (p1);
	\node[conn,above of=mul1] (m0) {};
	\draw[->] (m0) -- (mul1);
	\draw[->] (lcc) -- ++(0,10pt) -- ++(80pt,0) -- ++(0,-10pt) -- (lco);
	\node[conn,left of=D, xshift=10pt,yshift=-10pt] (uc) {};
	\node[conn,above of=uc, yshift=30pt] (c0) {};
	\draw[->] (c0) -- ++(55pt,0) to[vhpath] (mux1.north east);
	\node[above of=uc, yshift=61pt] (ui) {$M_{0,j}$};
	\draw[-] (ui) -- (uc);
	\node[below of=uc, yshift=-15pt] (uo) {$M_{0,j}$};
	\draw[->] (uc) -- (uo);
	\draw[->] (uc) -- (mul0);
	\draw[->] (lcc) to[vhpath] (mul0);

	\node[above of=inv, yshift=0pt] (mi0) {\tiny$1$};
	\node[above of=r0, yshift=-7pt] (mi1) {\tiny$0$};
	\end{tikzpicture}
};
\end{tikzpicture}
	\caption{The processing element $\GE_{i,\, j}$}
	\label{fig:ge}
\end{figure}
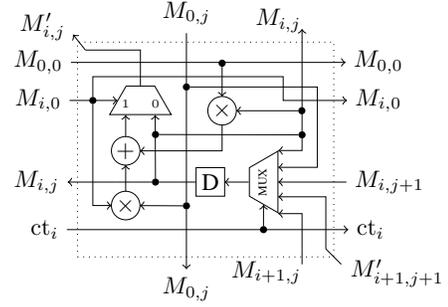
\subsection{Gabidulin's Algorithm}\label{sec:gabidulin}
In Algorithm~\ref{alg:gabidulin}, the matrix is first reduced to a triangular form.
It takes advantage of the property of the matrix so that it requires no division in the first stage.
In the first stage, we need to perform elimination on only one row.
We use a similar pivoting scheme like Algorithm~\ref{alg:ge}.
When a row is reduced to have only one non-zero element, a division is used to obtain one coefficient of $\bm{X}$.
Then it performs a backward elimination after getting each coefficient.
Hence we introduce a backward pivoting scheme, where the pivot element is always at the bottom-right corner.

In Algorithm~\ref{alg:gabidulin}, there are two $\tau \times \tau$ matrices over $\mathbb{F}_{q^m}$, $\bm{A}$ and $\bm{Q}$.
In step~\ref{alg:gabidulin}.2, it requires only $Q_{i,0}$'s to compute the coefficients.
To compute $Q_{i,0}$ in \eqref{eqn:gabiq}, it requires only $Q_{i-1,0}$ and $Q_{i-1,1}$.
And for $Q_{i,j}$ in \eqref{eqn:gabiq}, it requires only $Q_{i-1,j}$ and $Q_{i-1,j+1}$.
Recursively, only those $Q_{i,j}$'s where $i +j < \tau$ are necessary.
Actually, given any $i$, entries $Q_{i,0}, Q_{i+1,0}, \dotsc, Q_{\tau-1,0}$ can be computed with the entries $Q_{i-1,0}, Q_{i-1,1}, \dotsc, Q_{i-1,\tau-i}$.
With $Q_{0,0}, Q_{1,0}, \dotsc, Q_{i-2,0}$, we need to store only $\tau$ values to keep track of $\bm{Q}$.
Hence we reduce the storage of $\bm{Q}$ from $\tau\times \tau$ $m$-bit registers down to $\tau$.
We cannot reduce the storage of $\bm{A}$ to $\tau(\tau+1)/2$ because we have to use the pivoting scheme for short critical paths.

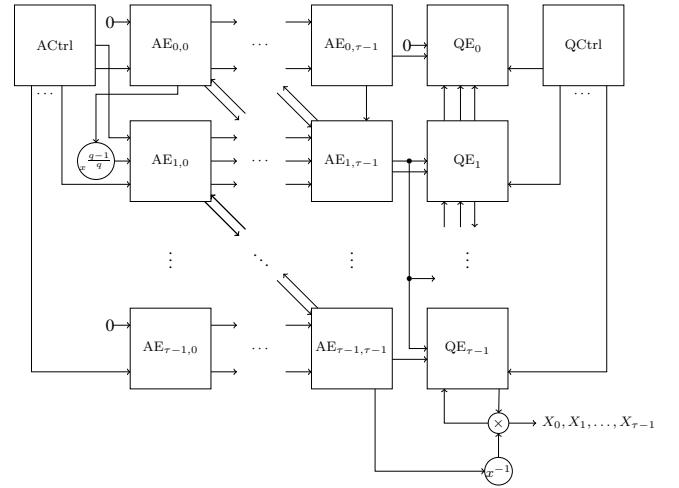
\begin{figure}[htbp]
	\centering
	\begin{tikzpicture}
\ifCLASSOPTIONonecolumn
		\node[scale=0.8]{
		\else
		\node[scale=0.65]{
		\fi
\begin{tikzpicture}[font=\fontsize{8pt}{8pt}]
	\matrix[row sep=20pt, column sep=20pt] {
	\node[peg] (actrl) {$\ACtrl$}; &\node[peg] (ae00) {$\AAE_{0,0}$}; & 
	\node{$\dotsb$};& \node[peg] (ae0n1) {$\AAE_{0,\tau-1}$};& \node[peg] (qe0) {$\QE_0$}; & \node[peg] (qctrl) {$\QCtrl$};\\
	& \node[peg] (ae10) {$\AAE_{1,0}$};& 
	\node {$\dotsb$}; & \node[peg] (ae1n1) {$\AAE_{1,\tau-1}$};& \node[peg] (qe1) {$\QE_1$};\\
	& \node {$\vdots$}; &
	\node{$\ddots$}; & \node {$\vdots$}; & \node {$\vdots$};\\
	& \node[peg] (aen10) {$\AAE_{\tau-1,0}$}; &
	\node {$\dotsb$}; & \node[peg] (aen1n1) {$\AAE_{\tau-1,\tau-1}$};& \node[peg] (qen1) {$\QE_{\tau-1}$};\\
	};

	\draw[->] (actrl.0) -- ++(8pt,0) to[vhpath] (ae10.150);
	\draw[->] (ae1n1.0) -- (qe1.180);
	\draw[->] (ae0n1.-15) -- (qe0.-165);
	\draw[->] (ae1n1.-15) -- (qe1.-165);
	\draw[->] (aen1n1.-15) -- (qen1.-165);

	\node[op,left of=ae10,xshift=-15pt,yshift=0pt] (a00) {$\scriptscriptstyle x^\frac{q-1}{q}$};
	\node[op, below of=qen1,xshift=18pt,yshift=-43pt] (inv) {$x^{-1}$};
	\node[op, below of=qen1,xshift=18pt,yshift=-15pt] (mul) {$\times$};
	\draw[->] (mul) to[hvpath] (qen1.-120);
	\draw[->] (qen1.-52) -- (mul);
	\draw[->] (inv) -- (mul);
	\draw[->] (aen1n1.-60) to[vhpath] (inv);
	\node[right of=mul, xshift=30pt] (xo) {$X_0,X_1,\dotsc,X_{\tau-1}$};
	\draw[->] (mul) -- (xo);
	\draw[->] (qctrl.-150) -- (qe0.-30);
	\draw[->] (qctrl.-120) to[vhpath] (qe1.-30);
	\node[below of=qctrl] {$\dotsb$};
	\draw[->] (qctrl.-60) to[vhpath] (qen1.-30);
	\draw[->] (actrl.-30) -- (ae00.-150);
	\draw[->] (actrl.-80) to[vhpath] (ae10.-150);
	\node[below of=actrl,xshift=-5pt] {$\dotsb$};
	\draw[->] (actrl.-120) to[vhpath] (aen10.-150);
	\draw[->] (ae00.-80) -- ++(0,-5pt) to[hvpath] (a00);
	\draw[->] (a00) -- ++(20pt,0);
	\draw[->] (ae0n1.-70) -- (ae1n1.70);
	\draw[->] (ae10.-50) -- ++(20pt,-20pt);
	\draw[<-] (ae10.-40) -- ++(20pt,-20pt);
	\draw[->] (ae00.-50) -- ++(20pt,-20pt);
	\draw[<-] (ae00.-40) -- ++(20pt,-20pt);
	\draw[->] (ae10.-50) -- ++(20pt,-20pt);
	\draw[<-] (ae10.-40) -- ++(20pt,-20pt);
	\draw[->] (ae1n1.130) -- ++(-20pt,20pt);
	\draw[<-] (ae1n1.140) -- ++(-20pt,20pt);
	\draw[->] (aen1n1.130) -- ++(-20pt,20pt);
	\draw[<-] (aen1n1.140) -- ++(-20pt,20pt);
	\draw[->] (qe1.120) -- (qe0.-120);
	\draw[->] (qe1.100) -- (qe0.-100);
	\draw[->] (qe1.80) -- (qe0.-80);
	\draw[<-] (qe1.-120) -- ++(0,-15pt);
	\draw[<-] (qe1.-100) -- ++(0,-15pt);
	\draw[->] (qe1.-80) -- ++(0,-15pt);
	\draw[<-] (qe0.-180) -- ++(-10pt,0) node {0 \space};
	\draw[->] (ae10.30) -- ++(15pt,0);
	\draw[<-] (ae1n1.150) -- ++(-15pt,0);
	\node[conn, right of=ae1n1,xshift=5pt] (ac) {};
	\draw[->] (ac) to[vhpath] (qen1.180);
	\node[conn, below of=ac,yshift=-40pt] (ac1) {};
	\draw[->] (ac1) -- ++(15pt,0);

	\draw[->] (ae10.0) -- ++(15pt,0);
	\draw[<-] (ae1n1.180) -- ++(-15pt,0);
	\draw[<-] (ae00.150) -- ++(-10pt,0) node {0 \space};
	\draw[<-] (aen10.150) -- ++(-10pt,0) node {0 \space};
	\draw[->] (ae00.30) -- ++(15pt,0);
	\draw[<-] (ae0n1.150) -- ++(-15pt,0);
	\draw[->] (aen10.30) -- ++(15pt,0);
	\draw[<-] (aen1n1.150) -- ++(-15pt,0);

	\draw[->] (ae00.-30) -- ++(15pt,0);
	\draw[<-] (ae0n1.-150) -- ++(-15pt,0);
	\draw[->] (ae10.-30) -- ++(15pt,0);
	\draw[<-] (ae1n1.-150) -- ++(-15pt,0);
	\draw[->] (aen10.-30) -- ++(15pt,0);
	\draw[<-] (aen1n1.-150) -- ++(-15pt,0);
\end{tikzpicture}
};
\end{tikzpicture}
\caption{Our architecture of Gabidulin's algorithm}
\label{fig:qea}
\end{figure}
\begin{figure}[htbp]
	\centering
		\begin{tikzpicture}
\ifCLASSOPTIONonecolumn
	\node[scale=0.9]{
\else
	\node[scale=0.7]{
\fi
	\begin{tikzpicture}[font=\fontsize{11pt}{11pt}, node distance=20pt]
	\matrix[draw, dotted, row sep=20pt, column sep=12pt, inner sep=10pt] {
	& & \node[op] (sqrt) {$\scriptscriptstyle x^{\frac{1}{q}}$};&\node[conn] (jc) {};\\
	\node[op] (mul) {$\times$}; & \node[op] (add) {$+$};\\
	& \node[reg,yshift=13pt] (D) {D}; & & \node[conn,yshift=13pt] (dccc) {};\\
	};
	\node[smux, above of=add, xshift=6pt, yshift=-2pt] (omux) {};
	\node[wmux, left of=D, xshift=-3pt] (imux) {};
	\node[above of=jc,yshift=20pt] (avi) {$A_{i-1,j}$};
	\node[conn,below of=mul,yshift=8pt] (aqc) {};
	\node[left of=aqc,xshift=-25pt] (aqi) {$A^{(q-1)/q}_{i-1,i-1}$};
	\node[right of=aqc,xshift=90pt] (aqo) {$A^{(q-1)/q}_{i-1,i-1}$};
	\node[right of=D,xshift=55pt] (ao) {$A_{i,j}$};
	\node[point, left of=aqc, xshift=9pt,yshift=52pt] (adi) {};
	\node[conn, right of=D,xshift=-8pt] (dc) {};
	\node[point, below of=jc, xshift=5pt, yshift=-60pt] (abi) {};
	\node[above of=aqi, yshift=45pt] (adii) {$A'_{i-1,j-1}$};
	\node[above of=adii,xshift=5pt] (ado) {$A'_{i,j}$};
	\node[left of=omux,xshift=-50pt] (ici) {$\ctau_i$};
	\node[conn,left of=omux] (icc) {};
	\node[right of=omux,xshift=52pt] (ico) {$\ctau_i$};
	\node[below of=ici,yshift=-50pt] (oci) {$\ctal_i$};
	\node[below of=ico,yshift=-50pt] (oco) {$\ctal_i$};
	\node[conn,below of=imux] (occ) {};
	\node[below of=oco,xshift=10pt] (dci) {$A_{i+1,j+1}$};
	\node[below of=dci,xshift=-10pt,yshift=5pt] (dco) {$A_{i,j}$};

	\draw (imux.north west) ++(3pt,0) node {\tiny$0$};
	\draw (imux.south west) ++(3pt,0) node {\tiny$1$};

	\draw (omux.south east) ++(0,3pt) node {\tiny$0$};
	\draw (omux.south west) ++(0,3pt) node {\tiny$1$};

	\draw[-] (dci) -- (abi);
	\draw[->] (occ) -- (imux.south);
	\draw[->] (oci) -- (oco);
	\draw[->] (icc) -- ++(0,-9pt) -- ++(50pt,0) to[vhpath] (ico);
	\draw[->] (ici) -- (omux.west);
	\draw[->] (adi) to[vhpath] (imux.north west);
	\draw[-] (adi) -- (adii);
	\draw[->] (abi) -- ++(-87pt,0) to[vhpath] (imux.south west);
	\draw[->] (D) -- (ao);
	\draw[->] (aqc) -- (mul);
	\draw[->] (aqi) -- (aqo);
	\draw[->] (omux.north) -- ++(0,18pt) -- ++(-30pt,0) -- (ado);
	\draw[->] (dc) -- (omux.south east);
	\draw[->] (add) -- (omux.south west);
	\draw[->] (avi) to[vhpath] (sqrt);
	\draw[->] (jc) to[vhpath] (add);
	\draw[->] (sqrt) to[hvpath] (mul);
	\draw[->] (mul) -- (add);
	\draw[->] (imux) -- (D);
	\draw[->] (dccc) -- ++(0,-17pt) -- (dco);
\end{tikzpicture}
};
\end{tikzpicture}
\caption{The processing element $\AAE_{i,j}$}
\label{fig:ae}
\end{figure}

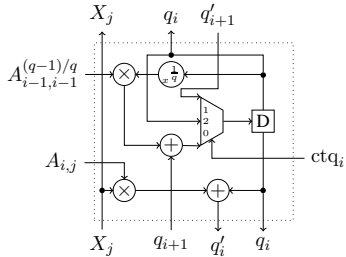
\begin{figure}[htbp]
	\centering
		\begin{tikzpicture}
\ifCLASSOPTIONonecolumn
	\node[scale=0.9]{
\else
	\node[scale=0.7]{
\fi
\begin{tikzpicture}[font=\fontsize{11pt}{11pt}, node distance=20pt]
	\matrix[draw, dotted, row sep=12pt, column sep=12pt, inner sep=10pt] {
	\node[op] (mul0) {$\times$};& \node[op] (sqrt) {$\scriptscriptstyle x^{\frac{1}{q}}$};\\
	& \node[op] (add0) {$+$}; & & \node[reg,yshift=13pt] (D1) {D};\\
	\node[op] (mul1) {$\times$}; & & \node[op] (add1) {$+$}; & \node[conn] (qoc) {};\\
	};
	
	\node[wmux, right of=add0, xshift=2pt, yshift=13pt] (mux) {};
	\node[conn, above of=D1, yshift=5pt] (qc) {};
	\node[above of=add0,yshift=49pt] (qi) {$q_i$};
	\draw[->] (mux) -- (D1);
	\draw[->] (D1) to[vhpath] (sqrt);
	\draw[->] (D1) to[vhpath] (add1);
	\draw[->] (add0) -- (mux.bottom right corner);
	\draw[->] (sqrt) -- (mul0);
	\draw[->] (mul0) to[vhpath] (add0);
	\draw[->] (mul1) -- (add1);
	\draw[->] (add1) -- ++(0,-22pt);
	\node[below of=add1, yshift=-10pt] (qo) {$q_i'$};
	\node[above of=add1, yshift=75pt] (qpi) {$q_{i+1}'$};
	\draw[->] (qpi) -- ++(0,-40pt) -- ++(-20pt,0) to[vhpath] (mux.bottom left corner);
	\node[conn, above of=sqrt, yshift=-9.5pt] (qcc) {};
	\draw[-] (qc) to[vhpath] (qcc);
	\draw[->] (qcc) -- (qi);
	\draw[->] (qcc) -- ++(-13pt,0) to[vhpath] (mux.west);
	\node[below of=add0, yshift=-33pt] (qp) {$q_{i+1}$};
	\draw[->] (qp) -- (add0);
	\node[left of=mul0, xshift=-24pt] (a0) {$A^{(q-1)/q}_{i-1,i-1}$};
	\node[left of=mul1, xshift=-14pt, yshift=15pt] (ai) {$A_{i,j}$};
	\draw[->] (ai) to[hvpath] (mul1);
	\draw[->] (a0) -- (mul0);
	\node[right of=D1, yshift=-20pt, xshift=15pt] (ci) {$\ctq_i$};
	\draw[->] (ci) to[hvpath] (mux.south);
	\node[below of=mul1,xshift=-12pt,yshift=-10pt] (xi) {$X_j$};
	\node[above of=mul0,xshift=-12pt,yshift=12pt] (xo) {$X_j$};
	\draw[->] (xi) -- (xo);
	\node[conn,left of=mul1,xshift=8pt] (xc) {};
	\draw[->] (xc) -- (mul1);

	\draw (mux.north west) ++(3pt,0) node {\tiny$1$};
	\draw (mux.south west) ++(3pt,0) node {\tiny$0$};
	\draw (mux.west) ++(3pt,0) node {\tiny$2$};

	\node[below of=qoc,yshift=-9pt] (qo) {${q_i}$};
	\draw[->] (qoc) -- (qo);
\end{tikzpicture}
};
\end{tikzpicture}
\caption{The processing element $\QE_{i}$}
\label{fig:qe}
\end{figure}

In our decoder, Algorithm~\ref{alg:gabidulin} is implemented by the regular architecture shown in Fig.~\ref{fig:qea}, which includes an array of $\tau\times\tau$ $\AAE$'s and a one-dimensional array of $\tau$ $\QE$'s.
The circuitry of the processing element $\AAE_{i,j}$ and $\QE_i$ is shown in Fig.~\ref{fig:ae} and \ref{fig:qe}.
The upper MUX in $\AAE$ controls the output sending upward along the diagonal.
Its control signal $\ctau_i$ is 1 for the second row and 0 for other rows since we update $\bm{A}$ one row in a cycle and we keep the pivot on the upper left corner in Step~\ref{alg:gabidulin}.1.
The control of the lower MUX in $\AAE$ is 0 for working on Step~\ref{alg:gabidulin}.1, and 1 for working on Step~\ref{alg:gabidulin}.2.
Similarly the control of the MUX in $\QE$ is 0 for working on Step~\ref{alg:gabidulin}.1, and 1 for working on Step~\ref{alg:gabidulin}.2.
But in Step~\ref{alg:gabidulin}.1, only part of $\QE$'s need update and others should maintain their values and their control signals $\ctq_i$'s are set to 2.
Initially, $A_{0,i} = E_i$ and $q_i=S_i$ for $i=0,1,\dotsc,\tau-1$.
Step~\ref{alg:gabidulin}.1 needs $\tau$ substeps. In the first $\tau-1$ substeps, $\ctal_{i+1}=0$, $\ctau_1=1$, $\ctq_0=\ctq_1=\dotsb=\ctq_i=2$, and $\ctq_{i+1}=\ctq_{i+2}=\dotsb=\ctq_{\tau-1}=0$ for substep~$i$.
In the last substep, $\ctau_1=0$ and all $\ctq_i$'s are set to 2.
This substep is to put the updated $\bm{A}$ into the original position.
In Step~\ref{alg:gabidulin}.2, the pivot is in the right lower corner, where we compute $X_i$'s.
Step~\ref{alg:gabidulin}.2 also needs $\tau$ substeps, in which all $\ctal_i$'s and $\ctq_i$'s are set to 1.
First $X_{\tau-1}$ is computed by $A_{\tau-1,\tau-1}^{-1} q_\tau-1$ where $q_{\tau-1}=Q_{\tau-1,0}$.
Note that the inversion may need $m-2$ clock cycles.
In each substep, the matrix $\bm{A}$ is moving down the diagonal so the $A_{i,i}$ to be inverted is always at the bottom right corner.
At the same time, the $q_i$'s are also moving down.
Basically, in substep~$p$, the architecture updates $q_i$'s to $Q_{i-p,0} - \sum_{j=\tau-1-p}^{\tau-1} A_{i,j}X_j$ for $i > p$ by doing one backward elimination at each substep.

\subsection{Low Complexity Linearized Interpolation}
It would seem that three registers are needed to store $F^{(i)}(x)$,  $w_j$'s, and  $\gamma^{(i)}_j$'s, respectively, in Algorithm~\ref{alg:minpoly}. However, we can implement Algorithm~\ref{alg:minpoly} with a single register of size $p+1$. First, we note that $w_j$'s are used to initialize $\gamma^{(0)}_j$'s, and only $\gamma^{(i)}_j$'s are used in the updates.
Second, after the $i$-th iteration of step~\ref{alg:minpoly}.2, the $q$-degree of $F^{(i+1)}(x)$ is no more than $i+1$ and we need only $\gamma^{(i+1)}_{i+1}, \gamma^{(i+1)}_{i+2}, \dotsc, \gamma^{(i+1)}_{p-1}$ thereafter. Thus, we can store the coefficients of $F^{(i+1)}(x)$ and $\gamma^{(i+1)}_{i+1}, \gamma^{(i+1)}_{i+2}, \dotsc, \gamma^{(i+1)}_{p-1}$ in a register of size $p+1$. We refer to this register as $\eta$ and index it $0, 1, \cdots, p$ from left to right. Note that $\gamma^{(i+1)}_{i+1}, \gamma^{(i+1)}_{i+2}, \dotsc, \gamma^{(i+1)}_{p-1}$ are stored at the lower end of the $\eta$ register, and the coefficients of $F^{(i+1)}(x)$ are stored at the higher end of the register. At each iteration, the content of the $\eta$ register is shifted to the left by one position, so that $\gamma^{(i)}_i$ is always stored at $\eta_0$.

\begin{algorithm}[Reformulated Algorithm for Minimal Linearized Polynomials]
	\SetKwFunction{rMP}{}

	\rMP

	\KwIn{Roots $w_0, w_1, \dotsc, w_{p-1}$}
	\KwOut{The minimal linearized polynomial $F(x)$}
	\BlankLine
	\renewcommand{\labelenumi}{\ref{alg:rminpoly}.\theenumi}
	\begin{enumerate}
		\item Initialization: $\eta_j^{(0)}=w_j$ for $j=0,1,\dotsc,p-1$, and $\eta_p^{(0)} = 1$.
		\item For $i=0, 1, \dotsc, p-1$,
			\begin{enumerate}
		\item If $\eta^{(i)}_0 \ne 0$,
			\begin{enumerate}
				\item For $j=0, 1, \dotsc, p-1-i$, $\eta_j^{(i+1)} = (\eta_{j+1}^{(i)})^{[1]} - (\eta_0^{(i)})^{q-1} \eta_{j+1}^{(i)}$;
				\item For $j=p-i, p-i+1, \dotsc, p$, $\eta_j^{(i+1)} = (\eta_j^{(i)})^{[1]} - (\eta_0^{(i)})^{q-1} \eta_{j+1}^{(i)}$;
			\end{enumerate}
		\item Otherwise, for $j=0, 1, \dotsc, p$, $\eta_j^{(i+1)} = \eta_{j+1}^{(i)}$.
		\end{enumerate}
		\item $F(x) = \sum_{i=0}^{p} \eta^{(p)}_i x^{[i]}$.
	\end{enumerate}
	\label{alg:rminpoly}
\end{algorithm}

\begin{figure}[htbp]
	\centering
		\begin{tikzpicture}
\ifCLASSOPTIONonecolumn
	\node[scale=0.9]{
\else
	\node[scale=0.7]{
\fi
\begin{tikzpicture}[
		hvpath/.style={to path={-| (\tikztotarget)}},
        vhpath/.style={to path={|- (\tikztotarget)}},
  		skip loop/.style={to path={- ++(0,#1) -| (\tikztotarget)}}]
	\matrix[row sep=10pt, column sep=20pt] {
	\node[pe] (pe0) {$\ME_0$}; & \node {$\dotsb$};&
	\node[pe] (pe7) {$\ME_{p}$};
	& \node[pe] (ctrl) {MCtrl};\\
	};

	\draw[->] (ctrl.140) to[vhpath] (pe7.40);
	\draw[<-] (pe7.-40) -- ++(10pt,0) node {$\,\,\tiny0$};
	\draw[<-] (pe7.0) -- ++(10pt,0) node {$\,\,\tiny0$};
	\draw[->] (ctrl.200) to[vhpath] (pe7.-20);
	\draw[->] (pe7.140) -- ++(-20pt,0);
	\draw[->] (pe7.180) -- ++(-20pt,0);
	\draw[->] (pe7.200) -- ++(-20pt,0);
	\draw[->] (pe7.220) -- ++(-20pt,0);
	\draw[->] (pe0.140) -- ++(-15pt,0) -- ++(0, 30pt) to[hvpath] (ctrl.90);

	\draw[->] (pe0.120) -- ++(0, 10pt);
	\draw (pe0.120) ++(0, 15pt) node{$F_{0}$};

	\draw[->] (pe7.120) -- ++(0, 10pt);
	\draw (pe7.120) ++(0, 15pt) node{$F_{p}$};

	\draw[<-] (pe0.40) -- ++(20pt,0);
	\draw[<-] (pe0.0) -- ++(20pt,0);
	\draw[<-] (pe0.-20) -- ++(20pt,0);
	\draw[<-] (pe0.-40) -- ++(20pt,0);
\end{tikzpicture}
};
\end{tikzpicture}
\caption{Architecture of linearized polynomial interpolation}
\label{fig:linear}
\end{figure}
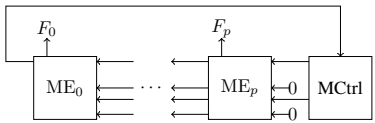

We note that the updates involve $\eta_{p+1}^{(i)}$, which is always set to zero (see Fig.~\ref{fig:linear}).
When an input $w_i$ is not linearly independent with $w_0,w_1,\dotsc,w_{i-1}$,  $\eta_0^{(i)}=0$. In this case, the algorithm simply ignores the input, and the $\eta_i$ registers are shifted to the left by one position. Hence, whether or not the inputs $w_0, w_1, \dotsc, w_{p-1}$ are linearly independent, the minimal linearized polynomial for the inputs will be available after $p$ iterations. This flexibility is important for our decoder architecture, since the number of linearly independent inputs varies.

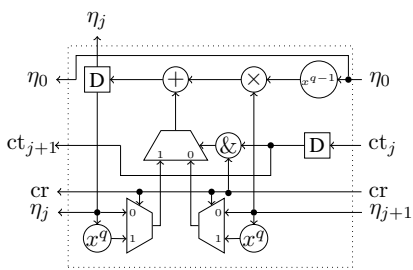
\begin{figure}[htbp]
	\centering
		\begin{tikzpicture}
\ifCLASSOPTIONonecolumn
	\node[scale=0.9]{
\else
	\node[scale=0.8]{
\fi
	\begin{tikzpicture}[font=\fontsize{11pt}{11pt}, node distance=20pt]
	\matrix[draw, dotted, row sep=15pt, column sep=15pt, inner sep=7pt] {
	\node[reg] (D) {D}; & \node[op] (add) {$+$}; & \node[op] (mul) {$\times$}; & \node[op] (power) {$\scriptscriptstyle x^{q-1}$};\\
	& \node[smux] (mux2) {};\\
	\\
	\node[op] (sq0) {$x^q$}; & & \node[op] (sq1) {$x^q$};\\
	};
	\node[op, right of=mux2, xshift=5pt] (and) {$\&$};
	\node[reg, right of=and, xshift=22pt] (DC) {D};
	\node[wmux, right of=sq0, yshift=6pt] (mux0) {};
	\node[emux, left of=sq1, xshift=0pt, yshift=6pt] (mux1) {};
	\node[conn, left of=DC, xshift=-2pt] (ct0) {};
	\node[conn, above of=sq0, yshift=-7.75pt] (rj0) {};
	\node[conn, above of=sq1, yshift=-7.75pt] (rjp0) {};
	\node[conn, above of=mux0, yshift=-4pt] (cr0) {};
	\node[conn, above of=mux1, yshift=-4pt] (cr1) {};
	\node[right of=rjp0, xshift=45pt] (rjp) {$\eta_{j+1}$};
	\node[right of=DC, xshift=11pt] (cti) {$\ct_j$};
	\node[left of=cti, xshift=-142pt] (ctip) {$\ct_{j+1}\,\,\,\,$};
	\node[point, right of=ctip, xshift=18pt] (ctip0) {};
	\node[right of=power, xshift=9pt] (r0) {$\eta_0$};
	\node[conn, right of=power, xshift=-6pt] (r00) {};
	\node[point, left of=D, xshift=10pt] (r0p0) {};
	\node[left of=r0p0, xshift=2pt] (r0p) {$\eta_0$};
	\node[left of=rj0, xshift=-7pt] (rj) {$\eta_j$};
	\node[right of=cr1, xshift=59pt] (cr) {$\crr$};
	\node[left of=cr0, xshift=-27pt] (crp) {$\crr$};
	\draw[->] (power) -- (mul);
	\draw[->] (rjp0) -- (mul);
	\draw[->] (rjp) -- (mux1.north east);
	\draw[->] (rjp0) -- (sq1);
	\draw[->] (rj0) -- (mux0.north west);
	\draw[->] (cr) -- (crp);
	\draw[->] (cr0) -- (mux0.north);
	\draw[->] (cr1) -- (mux1.north);
	\node[conn, below of=and, yshift=-2.5pt] (cr2) {};
	\draw[->] (cr2) -- (and);
	\draw[->] (D) -- (sq0);
	\draw[->] (DC) -- (and);
	\draw[->] (mux2.north) to[hvpath, skip loop=1mm] (add);
	\draw[->] (mux0.east) to[hvpath] (mux2.south west);
	\draw[->] (mux1.west) to[hvpath] (mux2.south east);
	\draw[->] (and) -- (mux2.east);
	\draw[->] (cti) -- (DC);
	\draw[->] (add) -- (D);
	\draw[->] (mul) -- (add);
	\draw[->] (r0) -- (power);
	\draw[-] (r00) to[vhpath, skip loop=4mm] (r0p0);
	\draw[->] (r0p0) -- (r0p);
	\draw[->] (rj0) -- (rj);
	\draw[->] (sq0) -- (mux0.south west);
	\draw[->] (sq1) -- (mux1.south east);
	\draw[-] (ct0) to[hvpath, skip loop=-5mm] (ctip0);
	\draw[->] (ctip0) -- (ctip) -- ++ (7pt,0);
	\node[above of=D, yshift=8pt] (rjp2) {$\eta_j$};
	\draw[->] (D) -- (rjp2);
	\draw (mux0.north west) ++(3pt,0) node {\tiny$0$};
	\draw (mux0.south west) ++(3pt,0) node {\tiny$1$};
	\draw (mux2.south west) ++(0,3pt) node {\tiny$1$};
	\draw (mux2.south east) ++(0,3pt) node {\tiny$0$};
	\draw (mux1.north east) ++(-3pt,0) node {\tiny$0$};
	\draw (mux1.south east) ++(-3pt,0) node {\tiny$1$};
\end{tikzpicture}
};
\end{tikzpicture}
\caption{The processing element $\ME_j$ ($x^q$ is a cyclic shift, and requires no hardware but wiring). For simplicity, we have omitted the superscripts of $\eta_j$}
\label{fig:me}
\end{figure}

Algorithm~\ref{alg:rminpoly} is implemented by the systolic architecture shown in Fig.~\ref{fig:linear}, which consists of $p+1$ processing elements ($\ME$'s).
The circuitry of the processing element $\ME_j$ is shown in Fig.~\ref{fig:me}.
The $\crr$ signal is 1 only when $\gamma_0 \ne 0$.
The $\ct_j$ signal for each cell is 1 only if $j < p - i$.
Basically, $\ct_j$ controls whether the update is for $F^{(i+1)}(x)$ or $\gamma^{(i+1)}_{i+1}, \gamma^{(i+1)}_{i+2}, \dotsc, \gamma^{(i+1)}_{p-1}$ as in Algorithm~\ref{alg:minpoly}.

\subsection{Decoding Failure}
A complete decoder declares decoding failure when no valid codeword is found within the decoding radius of the received word.
To the best of our knowledge, decoding failures of Gabidulin and KK codes were not discussed in previous works.
Similar to RS decoding algorithms, a rank decoder can return decoding failure when the roots of the error span polynomial $\lambda(x)$ are not unique.
That is, the root space of $\lambda(x)$ has a dimension smaller than the $q$-degree of $\lambda(x)$.
Note that this applies to both Gabidulin and KK decoders.
For KK decoders, another condition of decoding failure is when the total number of erasures and deviations exceeds the decoding bound $d-1$.

\subsection{Latency and Throughput}

We analyze the worst-case decoding latencies of our decoder architectures, in terms of clock cycles, in Table~\ref{tab:latency}.
\begin{table}[htbp]
	\centering
	\caption{Worst-case decoding latency (in terms of clock cycles). Gaussian elimination over $\mathbb{F}_{q^m}$ (root space in Gabidulin and KK decoders) has the longest critical path of one multiplier, one adder, one two-input MUX, and one five-input MUX.}
	\scalebox{0.83}{
	\begin{tabular}{|c|c|c|}
		\hline
		& Gabidulin & KK\\
		\hline
		$n$-RRE & - & $n(2N-n+1)/2$\\
		\hline
		Syndrome $\bm{S}$ & $n$ & $n$\\
		\hline
		$\lambda_U(x)$ & - & $2t$\\
		\hline
		$\sigma_D(x)$ & - & $2t$\\
		\hline
		$S_{DU}(x)$ & - & $2(d-1)$\\
		\hline
		BMA & $2t$ & $2t$\\
		\hline
		$S_{FD}(x)$ & - & $d-1$\\
		\hline
		$\bm{\beta}$ & - & $(m+2)(d-1)$\\
		\hline
		$\sigma_U(x)$ & - & $2t$\\
		\hline
		$\sigma(x)$ & - & $d-1$\\
		\hline
		root space basis $\bm{E}$ & $m(m+1)/2$ & $m(m+1)/2$\\
		\hline
		error locator $\bm{L}$ & $2t+mt$ & $(m+2)(d-1)$\\
		\hline
		error word $\bm{e}$ & $t$ & $2t$\\
		\hline
	\end{tabular}
	}
	\label{tab:latency}
\end{table}

As in~\cite{Bogdanov06}, the latency of Gaussian elimination for the $n$-RRE form is at most $n(2N-n+1)/2$ cycles.
Similarly, the latency of finding the root space is at most $m(m+1)/2$.

For Gabidulin's algorithm, it needs one cycle per row for forward elimination and the same for backward elimination.
For each coefficient, it takes $m$ cycles to perform a division.
Hence it needs at most $2(d-1)+m(d-1)$ and $2(d-1)+m(d-1)$ for $\bm{\beta}$ and $\bm{L}$ respectively.
The latencies of finding the minimal linearized polynomials are determined by the number of registers, which is $2t$ to accommodate $\lambda_D(x)$, $\sigma_D(x)$, and $\sigma_U(x)$, whose degrees are $\mu$, $\delta$, and $\mu$, respectively.
The $2t$ syndromes can be computed by $2t$ sets of multiply-and-accumulators in $n$ cycles.
Note that the computations of $S(x)$, $\lambda_U(x)$, and $\sigma_D(x)$ can be done concurrently.
The latency of RiBMA is $2t$ for $2t$ iterations.
The latency of a symbolic product $a(x)\otimes b(x)$ is determined by the $q$-degree of $a(x)$.
When computing $S_{DU}(x)$, we are concerned about only the terms of $q$-degree less than $d-1$ because only those are meaningful for the key equation.
For computing $S_{FD}(x)$, the result of $\sigma_D(x)\otimes S(x)$ in $S_{DU}(x)$ can be reused, so it needs only one symbolic product.
In total, assuming $n=m$, the decoding latencies of our Gabidulin and KK decoders are $n(n+3)/2+(n+5)t$ and $n(N+2)+4(n+5)t$ cycles, respectively.

One assumption in our analysis is that the unit that computes $x^{q-1}$ in Figs.~\ref{fig:qe} and \ref{fig:me} is implemented with pure combinational logic, which leads to a long CPD for large $q$'s.
To achieve a short CPD for large $q$'s, it is necessary to pipeline the unit that computes $x^{q-1}$. There are two ways to pipeline it: $x^{q-1} = x\cdot x^2 \dotsb x^{q/2}$ that requires $\log_2{q}-1$ multiplications, or $x^{q-1}= x^q/x$ that requires $m$ multiplications for division.
To maintain a short CPD, $x^{q-1}$ needs to be implemented sequentially with one clock cycle for each multiplication.
Let $c_{qm} = \min\{\log_2{q}-1,m\}$ and it requires at most $2(c_{qm}+2)t$ clock cycles for getting minimal linearized polynomials $\lambda_U(x)$, $\sigma_D(x)$, and $\sigma_U(x)$.
Similarly, it requires at most $c_{qm}(d-1)$ more cycles to perform forward elimination in Gabidulin's algorithm for the error locator, and the latency of this step will be $(m+c_{qm}+2)(d-1)$ cycles.

\begin{figure}[htbp]
	\centering
	\begin{tikzpicture}
\ifCLASSOPTIONonecolumn
	\node{
\else
	\node[scale=0.66]{
\fi
	\begin{tikzpicture}[draw=black!50]
    \matrix[row sep=4mm, column sep=4mm] {
	& & & \node (rsy) [reg] {$\bm{r}_s$}; & & \node (rsm) [reg] {$\bm{r}_{\sigma}$};\\
	& & & & & \node (ssm) [reg] {$\bm{S}_{\sigma}$};\\
        \node (words) [nonterminal] {Received};&
        \node (p2) [conn] {}; & \node (synds) [terminal]    {Syndromes};&
        \node (p3) [conn] {}; & \node (bma) [terminal]    {BMA};&
        \node (p4) [point] {};\\
        \node (codes) [nonterminal] {Corrected};&
		\node [point] {}; &
        \node (es) [terminal] {Error}; &
		\node (p7) [point] {};&
		\node (gabidulin) [terminal] {Gabidulin's};&
		\node (p5) [conn] {};&
		\node (rs)   [terminal]  {Roots}; & \node (sp) [point] {}; & \node (rp) [point] {};\\
& & & \node (ex) [reg] {$\bm{E}_{X}$};\\
	& & & & & \node (se) [reg] {$\bm{S}_{E}$};\\
	& & & \node (rx) [reg] {$\bm{r}_{X}$}; & & \node (re) [reg] {$\bm{r}_{E}$};\\
    };

	\node[below of=p2, yshift=20pt] {$\bm{r}$};
	\node[below of=p3, yshift=20pt] {$\bm{S}$};
	\node[above of=p4, yshift=-20pt] {$\sigma(x)$};
	\node[above of=p5, yshift=-20pt] {$\bm{E}$};
	\node[above of=p7, yshift=-20pt] {$\bm{X}$};
    { [start chain]
        \chainin (words);
		{[start branch=r]
		\chainin (rsy) [join=with p2 by {vhpath, tip}];
		\chainin (rsm) [join=by tip];
		\chainin (rp) [join=by hvpath];
		\chainin (re) [join=by {vhpath,tip}];
		\chainin (rx) [join=by tip];
		}
        \chainin (synds)    [join=by tip];
		{[start branch=s]
		\chainin (ssm) [join=with p3 by {vhpath, tip}];
		\chainin (sp) [join=by hvpath];
		\chainin (se) [join=by {vhpath,tip}];
		}
        \chainin (bma)   [join=by tip];
		\chainin (rs)    [join=by {hvpath, tip}];
		\chainin (gabidulin) [join=by tip,join=with se by {hvpath,tip}];
		{[start branch=e]
		\chainin (ex) [join=with p5 by {vhpath,tip}];
		}
		\chainin (es)    [join=by tip,join=with rx by {hvpath, tip},join=with ex by {hvpath,tip}];
		\chainin (codes) [join=by tip];
  }
\end{tikzpicture}
};
\end{tikzpicture}
\caption{Data flow of our pipelined Gabidulin decoder}
\label{fig:gpipeline}
\end{figure}
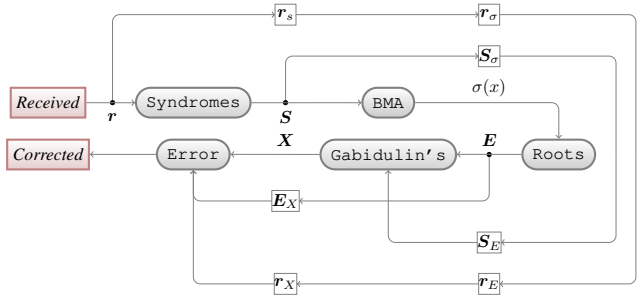

In our architectures, we use a block-level pipeline scheme for high throughput.
Data transfers between modules are buffered into multiple stages so the throughput is determined by only the longest latency of a single module.
For brevity, we present only the data flow of our pipelined Gabidulin decoder in Fig.~\ref{fig:gpipeline}.
The data in different pipeline stages are for different decoding sessions.
Hence these five units can work on five different sessions currently for higher throughput.
If some block finishes before others, it cannot start another session until all are finished.
So the throughput of our block-level pipeline decoders is determined by the block with the longest latency.
For Gabidulin decoders, the block of finding root space is the bottleneck that requires $m(m+1)/2$ cycles, the longest latency in the worst case scenario.
For KK decoders, the bottleneck is the RRE block, which requires $n(2N-n+1)/2$ cycles.

\section{Implementation Results and Discussions}\label{sec:impl}
To evaluate the performance of our decoder architectures, we implement our architectures for Gabidulin and KK codes for RLNC over $\mathbb{F}_{2}$. Note that although the random linear combinations are carried out over $\mathbb{F}_{2}$, decoding of Gabidulin and KK codes are performed over extension fields of $\mathbb{F}_{2}$.

We restrict $N$, the number of received packets, to save hardware while maintaining the error correction capability. We note that a large $N$ leads to more rows in the architecture in Fig.~\ref{fig:gaussian}. 
Note that we assume the input matrix is full rank as \cite{Silva08}.
When $N \geq n+d$, the number of deviations $\delta=N-n$ is at least $d$ and it is uncorrectable.
Hence in our implementation of KK decoders, we assume $N$ is less than $n+d$.

\subsection{Implementation Results}
We implement our decoder architecture in Verilog for an $(8,4)$ Gabidulin code over $\mathbb{F}_{2^8}$ and a $(16, 8)$ one over $\mathbb{F}_{2^{16}}$, which can correct errors of rank up to two and four, respectively.
We also implement our decoder architecture for their corresponding KK codes, which can correct $\epsilon$ errors, $\mu$ erasures, and $\delta$ deviations as long as $2\epsilon + \mu + \delta$ is no more than five or nine, respectively.
Our designs are synthesized using Cadence RTL Compiler~9.1 and FreePDK \SI{45}{\nm} standard cell library~\cite{Stine07}.
The synthesis results are given in Table~\ref{tab:syn8}.
In these tables, the total area includes both cell area and estimated net area, the gate counts are in equivalent numbers of 2-input NAND gates, and the total power includes both leakage and estimated dynamic power.
All estimations are made by the synthesis tool.
The throughput is computed as
		$(n \times m \times R)/(\text{Latency}_\text{Bottleneck}\times \text{CPD})$.

To provide a reference for comparison, the gate count of our $(8,4)$ KK decoder is only 62\% to that of the $(255, 239)$ RS decoder over the same field $\mathbb{F}_{2^8}$ in~\cite{Lee03}, which is 115,500.
So for Gabidulin and KK codes over small fields, which have limited error-correcting capabilities, their hardware implementations are feasible.
The area and power of decoder architectures in Table~\ref{tab:syn8} appear affordable except for applications with very stringent area and power requirements.

\begin{table}[htbp]
	\centering
	\caption{Synthesis results of decoders for Gabidulin and KK codes}
	\begin{tabular}{|c|c|c|c|c|c|}
		\hline
\multicolumn{2}{|c|}{Finite fields} & \multicolumn{2}{|c|}{$\mathbb{F}_{2^8}$} & \multicolumn{2}{|c|}{$\mathbb{F}_{2^{16}}$}\\
		\hline
\multicolumn{2}{|c|}{Codes} & Gab. & KK & Gab. & KK\\ \hline
\multicolumn{2}{|c|}{$(n, k)$ or $(n, m)$} & $(8, 4)$ & $(4, 4)$ & $(16, 8)$ & $(8, 8)$\\ \hline
		\multicolumn{2}{|c|}{Gates} & 18465 & 71134 & 116413 & 421477 \\
		\hline
		\multirow{3}{*}{Area (\si{\mm\squared})} & Cell & 0.035 & 0.133 & 0.219 & 0.791\\
		\cline{2-6}
		& Net & 0.053 & 0.202 & 0.320 & 1.163\\
		\cline{2-6}
		 & Total & 0.088 & 0.335 & 0.539& 1.954\\
		\hline
		\multicolumn{2}{|c|}{CPD (\si{\ns})} & 2.309 & 2.199  & 3.490 & 3.617\\
		\hline
		\multirow{2}{*}{Estimated} & Leakage & 0.281 & 1.084 & 1.690 & 6.216\\
		\cline{2-6}
		\multirow{2}{*}{Power (\si{\milli\watt})}& Dynamic & 14.205 & 54.106 & 97.905 & 313.065\\
		\cline{2-6}
		& Total & 14.486 & 55.190 & 99.595 & 319.281\\
		\hline
		\multicolumn{2}{|c|}{Latency (cycles)} & 70 & 216 & 236 & 752\\
		\hline
		\multicolumn{2}{|c|}{Bottleneck (cycles)} & 36 & 68 & 136 & 264\\
		\hline
		\multicolumn{2}{|c|}{Throughput (\si{\mega\bit\per\second})} & 385 & 214 & 270 & 134\\
		\hline
	\end{tabular}
	\label{tab:syn8}
\end{table}

\subsection{Implementation Results of Long Codes}
Although the area and power shown in Table~\ref{tab:syn8} are affordable and high throughputs are achieved, the Gabidulin and KK codes used have very limited block lengths $8$ and $16$. For practical network applications, the packet size may be large~\cite{Chou03}. One approach to increase the block length of a constant-dimension code is to lift a Cartesian product of Gabidulin codes \cite{Silva08}. We also consider the hardware implementations for this case.
We assume a packet size of 512 bytes, and use a KK code that is based on Cartesian product of 511 length-8 Gabidulin codes.
As observed in Section~\ref{sec:rre}, the $n$-RRE form allows us to either decode this long KK code in a serial, partly parallel, or fully parallel fashion. For example, more decoder modules can be used to decode in parallel for higher throughput.
We list the gate counts and throughput of the serial and factor-7 parallel schemes based on the $(8,4)$ KK decoder and those of the serial and factor-5 parallel schemes based on the $(16,8)$ KK decoder in Table~\ref{tab:cartesian8}.

\begin{table}[htbp]
	\centering
	\caption{Performance of KK decoders for 512-byte packets}
	\begin{tabular}{|c|c|c|c|c|}\hline
        $(n, m)$ & \multicolumn{2}{c|}{$(4, 4)$} & \multicolumn{2}{c|}{$(8, 8)$} \\ \hline
		Decoder & Serial & 7-Parallel & Serial & 5-Parallel \\		\hline
		Gates & 71134 & 497938 & 421477 & 2107385\\
		\hline
		Area (\si{\mm\squared}) & 0.335 & 2.345  & 1.954 & 9.770\\
		\hline
		CPD (\si{\ns}) & \multicolumn{2}{c|}{2.199} & \multicolumn{2}{c|}{3.617}\\
		\hline
		Est. Power (\si{\milli\watt}) & 55.190 & 386.330 & 319.281 & 1596.405\\
		\hline
		Latency (cycles) & 34896 & 5112 & 67808 & 13952\\
		\hline
		Throughput (\si{\mega\bit\per\second}) & 214 & 1498 & 134 & 670\\
		\hline
	\end{tabular}
	\label{tab:cartesian8}
\end{table}

In Table~\ref{tab:cartesian8}, we simply use multiple KK decoders for parallel implementations.
Parallel KK decoders actually share the same $\bm{\hat{A}}$, $\bm{\hat{L}}$, $\bm{\hat{X}}$, and $\lambda_U(x)$.
Hence, some hardware can be also shared, such as the left part of Gaussian elimination for reduction in Fig.~\ref{fig:ge} and the interpolation block for $\lambda_U(x)$.
With the same latency, these hardware savings are roughly 7\% of one single KK decoder.

\subsection{Discussions}
Our implementation results above show that the hardware implementations of RLNC over small fields and with limited error control are quite feasible, unless there are very stringent area and power requirements.
However, small field sizes imply limited block length and limited error control.
As shown above, the block length of a constant-dimension code can be increased by lifting a Cartesian product of Gabidulin codes.
While this easily provides arbitrarily long block length, it does not address the limited error control associated with small field sizes.
For example, a Cartesian product of $(8,4)$ Gabidulin codes has the same error correction capability as the $(8,4)$ KK decoder, and their corresponding constant-dimension codes also have the same  error correction capability.
If we want to increase the error correction capabilities of both Gabidulin and KK codes, longer codes are needed and in turn larger fields are required.
A larger field size implies a higher complexity for finite field arithmetic, and longer codes with greater error correction capability also lead to higher complexity.
It remains to be seen whether the decoder architectures continue to be affordable for longer codes over larger fields, and this will be the subject of our future work.

\section{Conclusion}\label{sec:con}
This paper presents novel hardware architectures for Gabidulin and KK decoders.
Our work not only reduces the computational complexity for the decoder but also devises regular architectures suitable for hardware implementations.
Synthesis results using a standard cell library confirm that our designs achieve high speed and high throughput.

\section*{Acknowledgment}
The authors would like to thank Dr. D. Silva and Prof. F. R. Kschischang for valuable discussions, and thank reviewers for their constructive comments.

\bibliographystyle{IEEEtran}

\appendices
\section{Proof of Lemma~\ref{lem:reduction}}\label{sec:proof1}
\begin{IEEEproof} This follows the proof of \cite[Proposition~7]{Silva08} closely. Let the RRE and an $n$-RRE forms of $\bm{Y}$ be
$\RRE(\bm{Y})  = \bigl[\begin{smallmatrix} W & \bm{\tilde{r}}\\
	0 & \bm{\hat{E}}\end{smallmatrix}\bigr]$ and
$\bar{\bm{Y}}' = \bigl[\begin{smallmatrix} W' & \bm{\tilde{r}}'\\
	0 & \bm{\hat{E}}'\end{smallmatrix}\bigr]$.
Since
the RRE form of $\bm{\hat{A}}$ is unique, $\bm{W} = \bm{W}'$. Thus, $\mu=\mu'$ and $\delta=\delta'$.
In the proof of \cite[Proposition~7]{Silva08}, $\mathcal{U}$ is chosen based on $\bm{W}$. Thus, we choose $\mathcal{U}=\mathcal{U}'$.
Since $\bm{\hat{L}}$ is uniquely determined by $\bm{W}$ and $\bm{\hat{L}}'$ is by $\bm{W}'$, we also have $\bm{\hat{L}}=\bm{\hat{L}}'$. Finally, choosing $\bm{r}'= \bm{I}_{\mathcal{U}'^c} \bm{\tilde{r}}'$, the rest follows the same steps as in the proof of \cite[Proposition~7]{Silva08}.
\end{IEEEproof}
\section{Proof of Lemma~\ref{lem:equivalence}}\label{sec:proof2}
\begin{IEEEproof} This follows a similar approach as in~\cite[Appendix C]{Silva08}. We have
\begin{align}
	\rank\begin{bmatrix}\bm{X}\\ \bm{Y}
	\end{bmatrix} &= \rank\begin{bmatrix}\bm{I} & \bm{x}\\
		\bm{I} + \bm{\hat{L}}'\bm{I}_{\mathcal{U}'}^T & \bm{r}'\\
		\bm{0} & \bm{\hat{E}}'
	\end{bmatrix}\nonumber\\
& = \rank\begin{bmatrix} \bm{\hat{L}}'\bm{I}_{\mathcal{U}'}^T & \bm{r}' - \bm{x}\\
	\bm{I}_{\mathcal{U}'^c}^T(\bm{I} + \bm{\hat{L}}'\bm{I}_{\mathcal{U}'}^T) & \bm{I}_{\mathcal{U}'^c}^T \bm{r}'\\
	\bm{0} & \bm{\hat{E}}'
\end{bmatrix}\label{eqn:iuc}\\
&=\rank\begin{bmatrix} \bm{\hat{L}}'\bm{I}_{\mathcal{U}'}^T & \bm{r}' - \bm{x}\\
	\bm{0} & \bm{\hat{E}}'
\end{bmatrix} + \rank \begin{bmatrix}
	\bm{I}_{\mathcal{U}'^c}^T & \bm{I}_{\mathcal{U}'^c}^T \bm{x}\end{bmatrix}\label{eqn:iu}\\
	&= \rank \begin{bmatrix} \bm{\hat{L}}' & \bm{r}' - \bm{x}\\
	\bm{0} & \bm{\hat{E}}'\end{bmatrix} + n - \mu'\nonumber
\end{align}
where \eqref{eqn:iuc} follows from $\bm{I}_{\mathcal{U}'}^T[\bm{I} + \bm{\hat{L}}'\bm{I}_{\mathcal{U}'}^T \mid \bm{r}] = \bm{0}$ and \eqref{eqn:iu} follows from $\bm{I}_{\mathcal{U}'}^T\bm{I}_{\mathcal{U}'^c}=0$.
Since $\rank \bm{X} + \rank\bm{Y} = 2n - \mu' + \delta'$, the subspace distance is given by
	$d_S(\langle \bm{X}\rangle, \langle \bm{Y}\rangle) = 2\rank\bigl[\begin{smallmatrix}\bm{X}\\ \bm{Y}
	\end{smallmatrix}\bigr] - \rank \bm{X} - \rank \bm{Y}
	= 2\rank\bigl[\begin{smallmatrix}\bm{\hat{L}}' & \bm{r}'-\bm{x}\\
		\bm{0} & \bm{\hat{E}}'
\end{smallmatrix}\bigr] - \mu' - \delta'$.
\end{IEEEproof}

\end{document}